\documentclass[12pt, draftcls, peerreview, onecolumn]{IEEEtran}

\usepackage{xcolor}
\usepackage{soul}
\usepackage{scalerel}
\usepackage[T1]{fontenc}
\usepackage{cite}
\usepackage[T1]{fontenc}
\usepackage{amsmath}
\usepackage{amsthm}
\usepackage{amssymb}
\usepackage{algorithm}
\usepackage{algpseudocode}
\usepackage{subcaption}
\usepackage{graphicx}
\usepackage{epstopdf}
\usepackage{pgfplots}
\usepackage{mathtools}
\usepackage{setspace}
\usepackage{bm}
\pgfplotsset{compat=1.8}
\usepackage{tikz}
\usepgfplotslibrary{patchplots}
\usetikzlibrary{positioning, arrows, shapes}
\usetikzlibrary{shapes.geometric}
\usetikzlibrary{positioning}
\usetikzlibrary{patterns}
\usetikzlibrary{decorations}
\usetikzlibrary{spy}
\tikzset{
	font={\fontsize{10pt}{10}\selectfont}}

\newcommand*{\vertbar}{\rule[-0.5ex]{0.5pt}{2.5ex}}
\newtheorem{theorem}{Theorem}
\newtheorem{lemma}[theorem]{Lemma}

\pgfplotsset{compat=1.11,
	/pgfplots/ybar legend/.style={
		/pgfplots/legend image code/.code={%
			\draw[##1,/tikz/.cd,yshift=-0.25em]
			(0cm,0cm) rectangle (3pt,0.8em);},
	},
}

\pgfplotsset{compat=1.11,
	/pgfplots/xbar legend/.style={
		/pgfplots/legend image code/.code={%
			\draw[##1,/tikz/.cd,yshift=-0.15em]
			(0cm,0cm) rectangle (10pt,0.4em);},
	},
}

\ifCLASSINFOpdf
\else
\fi
\begin{document}
\tikzstyle{block} = [draw, fill=black!20, rectangle, 
minimum height=3em, minimum width=6em, text width=6em,
text centered]
\tikzstyle{whiteblock} = [draw=none, minimum height=3em, minimum width=5em, text width=6em,
text centered]
\tikzstyle{sum} = [draw, fill=blue!20, circle, node distance=1cm]
\tikzstyle{input} = [coordinate]
\tikzstyle{output} = [coordinate]
\tikzstyle{pinstyle} = [pin edge={to-,thin,black}]
\tikzstyle{branch}=[fill,shape=circle,minimum size=1pt,inner sep=0pt]
\title{Sparse Vector Coding for Ultra Reliable and Low Latency Communications in 5G Systems}
%
%
%

\author{Hyoungju Ji,~\IEEEmembership{Member,~IEEE}
        Sunho Park,~\IEEEmembership{Member,~IEEE,}
        and~Byonghyo Shim,~\IEEEmembership{Senior Member,~IEEE}
\thanks{H. Ji, S. Park, and B. Shim are with the Institute of New Media and Communications and the Department of Electrical and Computer Engineering, Seoul National University, Korea (e-mail: hyoungjuji@islab.snu.ac.kr, shpark@islab.snu.ac.kr, bshim@snu.ac.kr).
	
	A part of this work has been presented in Information Theory and Applications (ITA) Workshop, San Diego, USA, Feb. 11-16, 2018 \cite{ITA}.
	
	This work was sponsored by the National Research Foundation of Korea (NRF) grant funded by the Korean government(MSIP) (2016R1A2B3015576 and 2014R1A5A1011478).

}
}
\textcopyright 2018 IEEE. Personal use of this material is permitted. Permission from IEEE must be
obtained for all other uses, in any current or future media, including reprinting/republishing this
material for advertising or promotional purposes, creating new collective works, for resale or
redistribution to servers or lists, or reuse of any copyrighted component of this work in other
works.

\maketitle
\begin{abstract}
Ultra reliable and low latency communication (URLLC) is a newly introduced service category in 5G to support delay-sensitive applications. In order to support this new service category, 3rd Generation Partnership Project (3GPP) sets an aggressive requirement that a packet should be delivered with $10^{-5}$ packet error rate within 1 ms transmission period. Since the current wireless transmission scheme designed to maximize the coding gain by transmitting capacity achieving long codeblock is not relevant for this purpose, a new transmission scheme to support URLLC is required. In this paper, we propose a new approach to support the short packet transmission, called sparse vector coding (SVC). Key idea behind the proposed SVC technique is to transmit the information after the sparse vector transformation. By mapping the information into the position of nonzero elements and then transmitting it after the random spreading, we obtain an underdetermined sparse system for which the principle of compressed sensing can be applied. From the numerical evaluations and performance analysis, we demonstrate that the proposed SVC technique is very effective in URLLC transmission and outperforms the 4G LTE and LTE-Advanced scheme.

\end{abstract}
%
\IEEEpeerreviewmaketitle
\section{Introduction}
Ultra reliable and low latency communication (URLLC) is a newly introduced service category in 5G to support delay-sensitive applications such as the tactile internet, autonomous driving, factory automation, cyber-physical system, and remote robot surgery \cite{ITU-R}. In order to support this new service category, 3rd Generation Partnership Project (3GPP) sets an aggressive requirement that a packet should be delivered with $10^{\text{-}5}$ block error rate (BLER) within 1 ms period \cite{MIoT}. One notable observation in these applications is that the transmit information is control (command) type information (e.g., move left/right, start/stop, rotate/shift, and speed up/down) or sensing information (e.g., temperature, moisture, pressure, and gas density) so that the amount of information to be delivered is tiny \cite{SI}. Since the current wireless transmission strategy designed to maximize the coding gain by transmitting capacity achieving long codeblock is not relevant to these URLLC scenarios, entirely new transmission strategy to support the short packet transmission is required. 
While there have been some efforts to improve the connection density, the medium access  latency, and the reliability of the re-transmission scheme for URLLC\cite{noma, urllc, short, DelayBudget, Div}, not much work has been done for the short-sized packet transmission except for the consideration of advanced channel coding schemes (e.g., polar code)~\cite{TR38913}.

In the current 4G systems, reliability of the data transmission is mainly achieved by the channel coding scheme \cite{lte}. Encoding at the basestation is done by the convolution coding and the decoding at the mobile terminal is done by the maximum likelihood decoding (MLD) or Turbo decoding. While this approach has shown to be effective in 4G systems, use of this scheme in URLLC scenario would be problematic since there is a stringent limitation on the packet length (and thus the parity size) yet the required reliability (target BLER = $10^{\text{-}5}$) is much higher than the current LTE-Advanced and LTE-Advanced Pro systems (target BLER = $10^{\text{-}2}{\sim}10^{\text{-}3}$) \cite{TR38913}. 
 
The purpose of this paper is to propose a new type of short packet transmission for URLLC that does not rely on the conventional channel coding principle. Key idea behind the proposed technique, henceforth referred to as {\it sparse vector coding} (SVC), is to transmit the short-sized information after the sparse vector transformation. To be specific, by mapping the information into the sparse vector and then transmitting it after the random spreading, we obtain an underdetermined sparse system for which the principle of compressed sensing can be applied \cite{tip}. It is now well-known from the theory of compressed sensing that an accurate recovery of a sparse vector is guaranteed with a relatively small number of measurements when the system matrix (a.k.a. sensing matrix) is generated at random \cite{Donoho}, which is achieved in our case via the random spreading. In fact, since the sparsity of the input vector is guaranteed by the sparse vector transformation, SVC decoding is achieved by the sparse signal recovery (more accurately, identification of nonzero positions in the transmit sparse vector). Therefore, the proposed scheme is very simple to implement and can be applied to wide variety of future wireless applications in which the amount of transmit information is sufficiently small. 

We note that there have been various efforts to use the support locations in the information encoding process \cite{PPM, Index, ImCode2, A2, AA}. For example, 
sparse mapping is conceptually similar to the position modulation (PM) and the index modulation (IM) techniques \cite{PPM, Index, ImCode2} in which the indices of the building block of the communication systems, such as pulses in optical systems, transmit antennas at the basestation or subcarrier groups in OFDM systems, are used to convey additional information bits. Also, in the single and multiple PM techniques, information is transmitted via the time sparsity  by using the combinations of the positions of optical pulses. In the spatial modulation-based IM technique, for example, additional information can be delivered by selectively using part of transmit antennas in the information transmission. Similar approach can also be found in Boolean multiple access channel \cite{A2, AA}. Our work is distinct from these studies in that we fully utilize the physical resources in the data transmission so that the loss, if any, caused by the underutilization of physical resources can be prevented. Also, in contrast to the IM technique where the receiver processing consists of two steps (the index recovery and symbol detection), decoding of the proposed SVC scheme is achieved by the identification of nonzero position called support identification \cite{tip}. The distinction of SVC over the conventional techniques is further strengthened by the fact that there is no random sensing mechanism (e.g., random spreading in SVC)  in the conventional schemes so that the compression of the transmit vector is not possible. 

From the performance analysis in terms of the decoding success probability and also numerical evaluations on the realistic setting, we demonstrate that the proposed SVC technique is very effective in short-size packet transmission and outperforms the 4G LTE and LTE-Advanced physical downlink control channel (PDCCH) scheme by a large margin in terms of reliability and transmission latency. 

The rest of this paper is organized as follows. In Section II, we briefly explain the short-sized packet transmission in 4G LTE and LTE-Advanced systems. In Section III, we present the proposed SVC scheme and explain the encoding and decoding operations. In Section IV, we analyze the success probability of SVC-encoded data transmission. Various implementation issues are discussed in Section V. In section VI, we present simulation results to verify the performance of the proposed scheme. We conclude the paper in Section VII.

\section{Short-sized Packet in LTE-Advanced Downlink}
In this section, we briefly review the control-type data transmission (PDCCH of 4G LTE systems) to illustrate the short-sized packet transmission in the conventional systems. PDCCH carries essential information for the mobile terminal when it tries to transmit or receive the data. To be specific, PDCCH carries small-sized information needed to decode the data channel (e.g., resource assignment, modulation order, code rate).  On top of these, cyclic redundancy check (CRC) is added to test the decoding error \cite{CRC}. Since the CRC bit stream is scrambled with a user index (called radio network temporary identifier), only the scheduled user can pass the CRC test.  

After the channel coding and symbol mapping,\footnote{e.g., convolution coding with rate $\frac{1}{3}$ and quadrature phase shift keying (QPSK) modulation are employed.} the modulated symbol vector $\mathbf{s}\in \mathbb{C}^{N \times 1}$ is transmitted. The corresponding received vector $\mathbf{y} \in \mathbb{C}^{m \times 1}$ is given by
\begin{equation}
\mathbf{y}=\mathbf{HR}\mathbf{s}+\mathbf{v},
\end{equation}
where $\mathbf{H}\in \mathbb{C}^{m \times m}$ is the diagonal matrix whose diagonal entry $h_{ii}$ is the channel component for each resource, $\mathbf{v} \sim \mathcal{C}\mathcal{N}(\mathbf{0}, \sigma^2_v\mathbf{I})$ is the additive Gaussian noise, and $\mathbf{R}\in \mathbb{C}^{m \times N}$ is the matrix describing the mapping between the symbol and resource element.
For example, when one symbol is mapped to a single resource, $\mathbf{R}$ would be the identity matrix ($\mathbf{R}=\mathbf{I}$). Whereas, if two resources are assigned to one symbol for the transmit diversity, then $\mathbf{R}$ would be $2N \times N$ matrix (e.g., if $N=2$, then $\mathbf{R}={\tiny \begin{bmatrix}
	1 & 0 & 1 & 0 \\ 0 & 1 & 0 & 1 \end{bmatrix}}^T$).

When one tries to improve the reliability with a small modification of current PDCCH, one can think of three options. 
The first option is to achieve the better coding gain by using lower code rate (i.e., $r=\frac{b}{2N}<r_{pdcch}=\frac{1}{3}$). This option is easy and straightforward but when the coded symbol length $N$ increases, transmission and processing latency will also increase, resulting in the violation of the URLLC requirement. The second option is to use the multiple resources to achieve the diversity gain ($m>N$). By combining multiple versions of the same symbol at the receiver, reliability of the symbol can be improved. The problem of this approach is that a large portion of wireless resources are consumed in achieving the diversity gain so that there would be a severe degradation of the resource utilization efficiency. The third option is to reduce the size of control information $b$. By removing some of the scheduling parameters, resources used for the control channel can be saved. Even in this case, it is not possible to remove essential information (e.g., CRC and user index) so that one cannot expect a dramatic reduction of control information. 

\section{Sparse Vector Coding}
\subsection{SVC Encoding and Transmission}
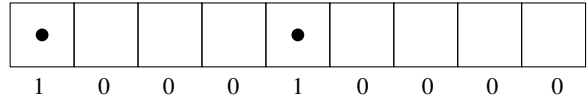
\begin{figure}
	\centering
		\begin{tikzpicture}
	[scale=0.85]
	\draw (0,0) -- (1,0) -- (1,1) -- (0,1) -- (0,0);
	\draw (0+1,0) -- (1+1,0) -- (1+1,1) -- (0+1,1) -- (0+1,0);
	\draw (0+2,0) -- (1+2,0) -- (1+2,1) -- (0+2,1) -- (0+2,0);
	\draw (0+3,0) -- (1+3,0) -- (1+3,1) -- (0+3,1) -- (0+3,0);
	\draw (0+4,0) -- (1+4,0) -- (1+4,1) -- (0+4,1) -- (0+4,0);
	\draw (0+5,0) -- (1+5,0) -- (1+5,1) -- (0+5,1) -- (0+5,0);
	\draw (0+6,0) -- (1+6,0) -- (1+6,1) -- (0+6,1) -- (0+6,0);
	\draw (0+7,0) -- (1+7,0) -- (1+7,1) -- (0+7,1) -- (0+7,0);
	\draw (0+8,0) -- (1+8,0) -- (1+8,1) -- (0+8,1) -- (0+8,0);
	\fill[black] (0.5,0.5) circle (0.1cm);
	\fill[black] (0.5+4,0.5) circle (0.1cm);
	\node[text width=5cm] at (3.3, -0.3) 	{\footnotesize{1~~~~~~0~~~~~~0~~~~~~0~~~~~~1~~~~~~0~~~~~~0~~~~~~0~~~~~~0}};
	\end{tikzpicture}
	\caption{Metaphoric illustration of SVC encoding. Information is mapped into the position of a sparse vector.}
	\label{fig:table}
\end{figure}

The key idea of the proposed SVC technique is to map the information into the positions of a sparse vector $\mathbf{s}$. 
Figuratively speaking, SVC encoding can be thought as marking a few dots to the empty table. As illustrated in Fig. \ref{fig:table}, if we try to mark dots to two cells out of 9, then there would be $\binom {9}{2}=36$ choices in total. In general, when we choose $K$ out of $N$ symbol positions, we can encode $\lfloor \log_2 \binom {N}{K} \rfloor$ bits of information. Example of one-to-one mapping between the information bit stream $\mathbf{w}$ and transmit sparse vector $\mathbf{s}$ is (see example in Table I)
\begin{equation}
\underbrace{\begin{matrix}
	0~0~0~0~0 \\ 0~0~0~0~1 \\ 0~0~0~1~1 \\ \vdots \\ 1~1~1~1~1
	\end{matrix}}_{b\text{-bit information }\mathbf{w} (b=5) }
\begin{matrix}
\longleftrightarrow \\ \longleftrightarrow \\ \longleftrightarrow \\\vdots\\ \longleftrightarrow
\end{matrix} ~~~
\underbrace{\begin{matrix}
	0~0~0~0~0~0~0~1~1 \\ 0~0~0~0~0~0~1~0~1 \\ 0~0~0~0~0~1~0~0~1 \\\vdots \\ 1~0~0~0~0~0~0~0~1 
	\end{matrix}}_{K-\text{sparse vector} ~\mathbf{s} ~(K=2)}  \begin{matrix}
\\  \\  \\ \\ _{\textstyle .} 
\end{matrix}\nonumber 
\end{equation} 
 
{
\begin{table}[]\footnotesize
	\centering
	\caption{Example of mapping between the information $\mathbf{w}$ and the sparse vector $\mathbf{s}$}
	\label{my-label}
	\begin{tabular}{llll}
		\cline{1-1} \cline{3-4}
		\textbf{Input:}  			      & \multicolumn{1}{c|}{~~~~~~~~~~~~~~~~~~~~~} & \multicolumn{1}{c|}{(\textbf{w})$_{(10)}$} & \multicolumn{1}{c|}{\textbf{s}} \\ \cline{3-4} 
		~~Size of sparse vector $N$,      & \multicolumn{1}{l|}{~~~~~~} & \multicolumn{1}{c|}{0} & \multicolumn{1}{l|}{000011} \\ \cline{3-4} 
		~~information vector $\mathbf{w}$ & \multicolumn{1}{l|}{~~~~~~} & \multicolumn{1}{c|}{1} & \multicolumn{1}{l|}{000101} \\ \cline{3-4} 
		\textbf{Output:}                  & \multicolumn{1}{l|}{} & \multicolumn{1}{c|}{2} & \multicolumn{1}{l|}{000110} \\ \cline{3-4} 
		~~Sparse vector $\mathbf{s}$      & \multicolumn{1}{l|}{} & \multicolumn{1}{c|}{3} & \multicolumn{1}{l|}{001001} \\ \cline{1-1} \cline{3-4} 
		$a:=0$                            & \multicolumn{1}{l|}{} & \multicolumn{1}{c|}{4} & \multicolumn{1}{l|}{001010} \\ \cline{3-4} 
		\textbf{for} $i=2$ \textbf{to} $N$ \textbf{do}  & \multicolumn{1}{l|}{} & \multicolumn{1}{c|}{5} & \multicolumn{1}{l|}{001100} \\ \cline{3-4} 
		~\textbf{for} $j=1$ \textbf{to} $i-1$ \textbf{do}  & \multicolumn{1}{l|}{} & \multicolumn{1}{c|}{6} & \multicolumn{1}{l|}{010001} \\ \cline{3-4} 
		~~\textbf{if} $a = \left(\mathbf{w}\right)_{(10)}$  &  \multicolumn{1}{l|}{}   & \multicolumn{1}{c|}{7}   & \multicolumn{1}{l|}{010010}   \\ \cline{3-4} 
		~~~$\mathbf{s} := \left(2^{i}+2^{j} \right)_{(2)}$  & \multicolumn{1}{l|}{} & \multicolumn{1}{c|}{8} & \multicolumn{1}{l|}{010100} \\ \cline{3-4} 
		~~\textbf{end if}  & \multicolumn{1}{l|}{} & \multicolumn{1}{c|}{9} & \multicolumn{1}{l|}{011000} \\ \cline{3-4} 
		~~$a:=a+1$  & \multicolumn{1}{l|}{} & \multicolumn{1}{c|}{10} & \multicolumn{1}{l|}{100001} \\  \cline{3-4} 
		~\textbf{end for} &  &  & \text{:}  \\ 
		\textbf{end for} & &   &  \text{:} \\  \cline{1-1} 
		\multicolumn{4}{l}\textbf{Note:} $\left(\mathbf{w}\right)_{(10)}$ is decimal expression of binary vector $\mathbf{w}$ and $\left({w}\right)_{(2)}$\\
		\multicolumn{4}{l}{ ~~~~~~~is binary expression of integer ${w}$.} 
	\end{tabular}
\end{table}
}

After the sparse mapping, each nonzero element in $\mathbf{s}$ is spread into $m$ resources using the codeword (spreading sequence) in the spreading codebook $\mathbf{C}$. While it is possible to allocate resources either in time, frequency axis or hybrid of these, in this work, we assume that they are allocated in the frequency axis (see Fig. 2(a)). This choice will not affect the system model but minimizes the transmission latency. As a result of this spreading process called the {\it multi-code spreading}, the resource mapping matrix $\mathbf{R}$ in (1) is replaced with the codebook matrix $\mathbf{C} = [\mathbf{c}_1~\mathbf{c}_2~\cdots~\mathbf{c}_N]$ where {$\mathbf{c}_i=[c_{1i}~c_{2i}~\cdots~c_{mi}]^T$} is the spreading sequence. For example, if the first and the third element of $\mathbf{s}$ are nonzero, then the transmit vector after spreading is
\begin{eqnarray}
\label{eq:e3}
\mathbf{x} &=&\mathbf{Cs} \nonumber \\ 
&=& s_1 \mathbf{c}_1 + s_3 \mathbf{c}_3.
\end{eqnarray}
Since the positions of nonzero elements are chosen at random, the codebook matrix $\mathbf{C}$ should be designed such that the transmit vector $\mathbf{x}$ contains enough information to recover the sparse vector $\mathbf{s}$ irrespective of the selection of the nonzero positions. It has been shown that if entries of the codebook matrix $\mathbf{C}$ are generated at random, e.g., sampled from Gaussian or Bernoulli distribution, then an accurate recovery of the sparse vector is possible as long as $m=\mathcal{O}\left(K \log N \right)$~\cite{Donoho}. Example of $\mathbf{C}$ for $m=5$ and $N=10$, when elements of $\mathbf{c}_i$ are chosen from the Bernoulli distribution, is given by 
\begin{equation}
\label{eq:444}
\mathbf{C}
= \frac{1}{\alpha}\begin{bmatrix}
~~1 & ~~1 & ~~1 & ~~1 & -1 & ~~1 & -1 & ~~1& -1& -1\\
~~1 & -1 & ~~1 & -1 & ~~1 & -1 & ~~1 & -1& -1& ~~1\\
~~1 & ~~1 & -1 & -1 & ~~1 & ~~1 & -1 & -1& ~~1& ~~1\\
~~1 & -1 & -1 & ~~1 & ~~1 & -1 &  ~~1 & ~~1& -1& -1\\
-1 & ~~1 & ~~1 & ~~1 & -1 & -1 & -1 & -1& ~~1& ~~1\\
\end{bmatrix}_{\textstyle ,}
\end{equation}
where $\alpha$ is the normalization factor depending on the modulated symbols (see Section V.B). The corresponding received signal $\mathbf{y}$ is 
\begin{eqnarray}
\mathbf{y}  &=& \mathbf{Hx}+\mathbf{v} \nonumber \\ &=& 	 	{\begin{bmatrix}
	\mathbf{H}\mathbf{c}_{1}  &  \mathbf{H}\mathbf{c}_{3}
	\end{bmatrix}
}
{
	\begin{bmatrix}
	s_1 \\ s_{3}
	\end{bmatrix}
	+ \mathbf{v}}.
\end{eqnarray}
In general, the received  vector $\mathbf{y}$ is given by 
\begin{eqnarray}
\mathbf{y} &=& \mathbf{H}\mathbf{Cs} +\mathbf{v} \nonumber \\
&=&  
\begin{bmatrix}
h_{11} &    &  \\
& \ddots   &  \\
&  & h_{mm} \\
\end{bmatrix}
\begin{bmatrix}
\vertbar  &  & \vertbar\\
\mathbf{c}_{1} &  \dots & \mathbf{c}_{N}\\
\vertbar &   &\vertbar\\
\end{bmatrix}
\begin{bmatrix}
s_1\\
\vdots\\
s_N
\end{bmatrix}
+	\begin{bmatrix}
v_1\\
\vdots\\
v_m
\end{bmatrix}_{\textstyle .}
\end{eqnarray}
It is worth mentioning that an accurate recovery of the sparse vector $\mathbf{s}$ is unnecessary in SVC since the decoding of the information vector is achieved by the identification of nonzero positions, not the actual values of this vector. The fact that the decoding is done by the support\footnote{Support is the set of nonzero elements. For example, if $\mathbf{s}=[0~0~1~0~0~1]$, the $\Omega_\mathbf{s}=\{3, 6\}$.} identification greatly simplifies the decoding process and also reduces the chance of decoding failure. The overall structure of the proposed SVC is depicted in Fig. 2(b). 

\begin{figure}
	\centering
		\begin{tikzpicture}[auto, node distance=1.5cm,>=latex', scale=0.6, decoration={brace,amplitude=2pt}]
	\draw [->] (0,0) --(7,0)node[anchor=north] {\scriptsize{time}};
	\draw [->] (0,0) --(0,8)node[anchor=west] {\scriptsize{frequency}};
	
	\draw [->] (9,0) --(16,0) node[anchor=north] {\scriptsize{time}};
	\draw [->] (9,0) --(9,8) node[anchor=west] {\scriptsize{frequency}\scriptsize};
	
	\draw [line width= 1.5, blue] (1,1) -- (1, 1+0.4*15)-- (2.2,1+0.4*15) -- (2.2,1)--(1,1);

	\foreach \x in {0,...,6}
	  \foreach \y in {0,1,2,3,4,5,9,10,11,12,13,14}
	    	{\draw [-] (1+0.4*\x,1+0.4*\y)--(1+0.4*\x,1.4+0.4*\y)--(1.4+0.4*\x,1.4+0.4*\y)--(1.4+0.4*\x,1+0.4*\y)--(1+0.4*\x,1+0.4*\y);}
	    	
	\foreach \x in {0,...,6}
		\foreach \y in {6,...,8}
			{\draw [dotted] (1+0.4*\x,1+0.4*\y)--(1+0.4*\x,1.4+0.4*\y)--(1.4+0.4*\x,1.4+0.4*\y)--(1.4+0.4*\x,1+0.4*\y)--(1+0.4*\x,1+0.4*\y);}
    
    \foreach \x in {0, 3, 6}
    	\foreach \y in {0, 4, 9, 13}
    		{\filldraw [-, red, pattern=north east lines] (1+0.4*\x,1+0.4*\y)--(1+0.4*\x,1.4+0.4*\y)--(1.4+0.4*\x,1.4+0.4*\y)--(1.4+0.4*\x,1+0.4*\y)--(1+0.4*\x,1+0.4*\y);}
    		
    \draw [->, blue] (4.1,4.3) --(2.2, 4)node[anchor= west,  near start] {\scriptsize{~~PDCCH region}};
    
    \draw [->, red] (4.3,5.3) --(3.8, 5)node[anchor= west,  near start] {\scriptsize{Pilot symbol}};
    
    \foreach \x in {0,...,6}
    	\foreach \y in {0,1,2,3,4,5,9,10,11,12,13,14}
    		{\draw [-] (9+1+0.4*\x,1+0.4*\y)--(9+1+0.4*\x,1.4+0.4*\y)--(9+1.4+0.4*\x,1.4+0.4*\y)--(9+1.4+0.4*\x,1+0.4*\y)--(9+1+0.4*\x,1+0.4*\y);}
    
    \foreach \x in {0,...,6}
    	\foreach \y in {6,...,8}
    		{\draw [dotted] (9+1+0.4*\x,1+0.4*\y)--(9+1+0.4*\x,1.4+0.4*\y)--(9+1.4+0.4*\x,1.4+0.4*\y)--(9+1.4+0.4*\x,1+0.4*\y)--(9+1+0.4*\x,1+0.4*\y);}
    		
    \draw [line width= 1.0, red] (9+1, 1) -- (9+1, 1+0.4*15)-- (10+0.4, 1+0.4*15) -- (10+0.4, 1)--(9+1, 1);
    \foreach \y in {0,1,2,3,4,5,9,10,11,12,13,14}
    	{\filldraw [-, red, pattern=north east lines] (9+1,1+0.4*\y)--(10,1.4+0.4*\y)--(9+1.4,1.4+0.4*\y)--(9+1.4,1+0.4*\y)--(9+1,1+0.4*\y);}

        \foreach \y in {0,1,2,3,4,5,9,10,11,12,13,14}
    {\filldraw [-, fill=gray] (9+1+0.8,1+0.4*\y)--(10+0.8,1.4+0.4*\y)--(9+1.4+0.8,1.4+0.4*\y)--(9+1.4+0.8,1+0.4*\y)--(9+1+0.8,1+0.4*\y);}	
     \draw [line width= 1, black] (9+1+0.8, 1) -- (9+1+0.8, 1+0.4*15)-- (10+0.4+0.8, 1+0.4*15) -- (10+0.4+0.8, 1)--(9+1+0.8, 1);	
        
        \draw [line width= 1.5, blue] (9+1+0.4, 1) -- (9+1+0.4, 1+0.4*15)-- (10+0.4+0.4, 1+0.4*15) -- (10+0.4+0.4, 1)--(9+1+0.4, 1);	
        
        \draw [->, red] (11.5,7.8) --(10.2, 7)node[anchor= south west,  near start] {~\scriptsize{Pilot region}};
        \draw [->, blue] (12,7.5) --(10.5, 6.8)node[anchor= west,  near start] {~\scriptsize{Control region}};
        \draw [->, black] (13, 6.5) --(10.9, 6)node[anchor= west,  near start] {~~\scriptsize{Data region}};
        
        \draw [->] (0.5,0.7)--(1,0.7);
        \draw [->] (0.5+2.2,0.7)--(1+1.2,0.7)node[anchor= west,  near start]{~\scriptsize{Control transmission}};
        \draw [-] (1, 0.5)--(1, 0.85);
        \draw [-] (2.2, 0.2)--(2.2, 0.85);
        
        \draw [line width= 1, black] (2.2, 1) --(4,1);
         \draw [line width= 1, dotted] (4, 1) --(4.5,1);
  
          \draw [line width= 1, black] (2.2, 1+0.4*15) --(4,1+0.4*15);
  \draw [line width= 1, dotted] (4, 1+0.4*15) --(4.5,1+0.4*15);
         
        \draw [->] (2.2,0.3)--(2.8,0.3)node[anchor= west,  near end]{~\scriptsize{Data transmission}};

        \draw [->] (9+0.5,0.7)--(9+1,0.7);
        \draw [->] (9+0.5+2.2,0.7)--(9+1+1.2,0.7)node[anchor= west,  near start]{~\scriptsize{URLLC transmission}};
        
        \draw [decoration, decorate, color=blue] (9.8,1) -- (9.8,7)
        node [midway,anchor=east,inner sep=2pt, outer sep=1pt]{\tiny$m$};

	\end{tikzpicture}
	\\ \scriptsize{(a)}\\
	\begin{tikzpicture}[auto, node distance=1.8cm,>=latex', scale=1]
	\node [whiteblock, node distance=1.5cm, scale=0.9] (Quantized data) {Data information};
	\node [branch, node distance=0.8cm, right of=Quantized data] (b1) {};
	\node [block, right of=Quantized data,
	node distance=3.1cm, scale=0.9] (system) {Sparse mapping};
	\draw [-] (Quantized data) --  (b1);
	\draw [->] (Quantized data) -- node[name=u] {\scriptsize{$b$ bits}~~~~} (system);
	\node [block, right of=system, node distance=3.0cm, scale=0.9] (ss) {Multi-code spreading};
	\node [branch, node distance=2.6cm, scale=0.9, right of=ss] (cb) {};
	\node [whiteblock, node distance=1.5cm, scale=0.9, below of=cb] (textchannel) {\scriptsize{Channel}};
	\node [cloud,cloud puff arc=150, aspect=2, inner ysep=1em, draw, node distance=1.35cm, below of=cb, scale=0.65] (channel) {~~~~~~~~~~};
	\node [branch, node distance=1.83cm, scale=0.9, below of=textchannel] (yy) {};
	\node [block, below of=ss, node distance=3cm, scale=0.9] (SI) {Support detection};
	\node [block, below of=system, node distance=3cm, scale=0.9] (Re) {Sparse demapping};
	\draw [->] (system) -- node[name=s] {\tiny{$N$ symbols}} (ss);
	\node [whiteblock, node distance=0.5cm, scale=0.9, below of=s] (sss) {$\mathbf{{s}}$};

	\node [whiteblock, below of=Quantized data, node distance=3cm, scale=0.9] (out) {Decoded information};
	\draw [->] (Re) -- node[name=ihat] {} (out);
	
	\draw [-] (ss) -- node [name=x, near start] {\scriptsize{~~~~~~~~~~~$\mathbf{x=Cs}$}}(cb);
	\draw [->] (cb) -- (textchannel);
	\draw [-] (textchannel) -- (yy);
	\draw [->] (yy) -- node [name=y, near end] {~~~~~$\mathbf{y}$}(SI);
	\draw [->] (SI) -- node [name=ohat] {$\hat{\mathbf{s}}$}(Re);
	
	\node[text=blue, fill=white, above left= 1mm and -7mm of ss] (trans) {\scriptsize{Basestation encoding}};
	\draw[blue] (trans.west)-|([xshift=-4mm]system.west)|-([yshift=-3mm]system.south)-|([xshift=3mm]ss.east)|-(trans.east);
	
	\node[text=blue, fill=white, above right= 2mm and -40mm of SI] (reci) {\scriptsize{Decoding at mobile station}};
	\draw[blue] (reci.west)-|([xshift=-4mm]Re.west)|-([yshift=-3mm]Re.south)-|([xshift=3mm]SI.east)|-(reci.east);
	\end{tikzpicture}
	\\ \scriptsize{(b)}
	\label{fig:sm}
	\caption{SVC-based packet transmission: (a) packet structure of 4G (left) and the URLLC packet (right) and (b) the block diagram for the proposed SVC technique.}
	\label{fig:system}
\end{figure}
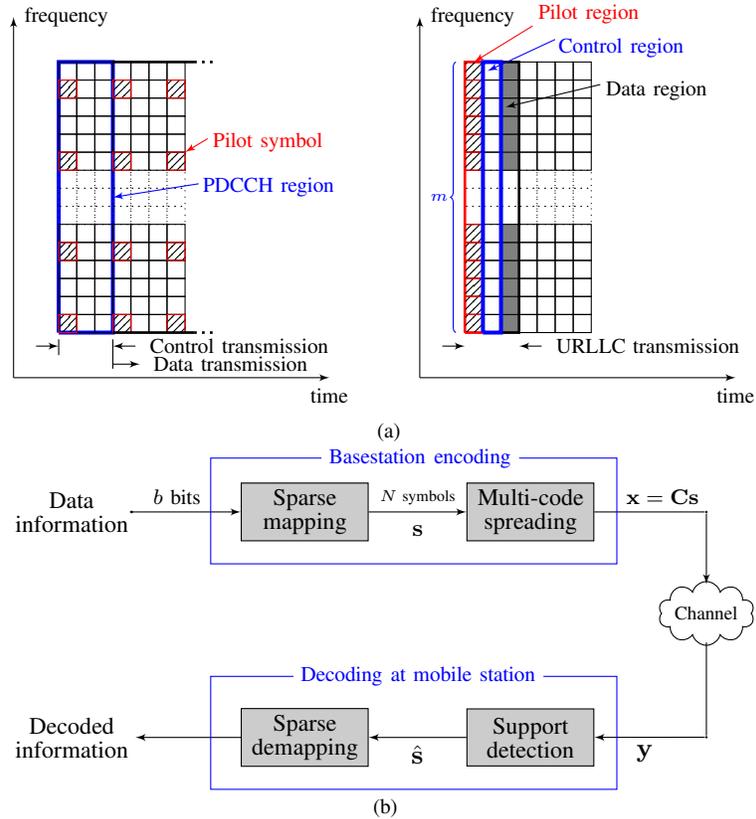
 The benefits of SVC can be summarized as follows; 
 First, the transmission power of the data channel is concentrated on the nonzero elements of an information vector. Thus, when compared to the conventional system in which the transmission power is uniformly distributed across all symbols, effective transmit power per symbol is higher. Second, the SVC decoding process achieved by the sparse recovery algorithm lends itself to the test of decoding success/failure so that the CRC operation is unnecessary. This directly implies that the code rate of SVC can be made smaller than the rate of PDCCH. Specifically, when the number of resources used for the data channel is $m$ and the QPSK modulation is used, the code rate of SVC is $r_{svc}=\frac{b_{i}}{2m}$ ($b_i$ is the number of information bits) and the code rate of PDCCH is $r_{pdcch}=\frac{(b_i + b_c)}{2m}\left(=\frac{1}{3}\right)$. If the number of CRC bits is $b_c=\beta b_i$ ($\beta>0$), then $m=\frac{3}{2}(b_i+b_c)= \frac{3}{2}(b_i+\beta b_i)$. Thus, the code rate of SVC can be expressed in term of $\beta$ as
    \begin{equation}
 	r_{svc}=\frac{b_{i}}{2m} = \frac{1}{{3(1 + \beta )}} < \frac{1}{3}=r_{pdcch}.
 	\end{equation}
 Third,  when $m$ is sufficiently large, the basestation can easily assign the distinct codebook $\mathbf{C}$ for each user. This is because codebook matrices can be made near orthogonal  by using a properly designed codebook generation mechanism.\footnote{The correlation between two distinct columns of  random matrix decreases exponentially as the dimension of a column increases (see, e.g., \cite[Theorem 1]{Cai}).} For example, when $m=42$ and the codebook is generated by the Bernoulli distribution, then there are $2^{42}$ different spreading sequences $\mathbf{c}_i$. Thus, if $N=96$, then the basestation can support maximally $2^{35} (\approx \frac{2^{42}}{96})$ devices. Last but not least important benefit of SVC is that the implementation cost is small and the processing latency is low. Encoding is done via a simple injective mapping and spreading, which can be easily realized by the look-up table and addition/subtraction operations and the decoding is performed by the support detection and demapping. In particular, since the sparsity $K$ is small and also known to the receiver, one can decode the SVC packet using a simple sparse recovery algorithm such as orthogonal matching pursuit (OMP) \cite{tip}.\footnote{Most of CS algorithm finds out the solution without the prior knowledge of the sparsity $K$. However, when $K$ is known in advance, one can recover the sparse vector more accurately by using the sparsity-aware recovery technique \cite{A1}.} Comparisons of PDCCH and SVC are summarized in Table. II.
\begin{table}[t!]\footnotesize
	\centering
	\caption{PDCCH versus SVC technique}
	\begin{tabular}{p{3cm}|p{4cm}|p{4cm}}
		\hline 
		& PDCCH & SVC technique \\ 
		\hline \hline
		Coding ~~~~~~~~~~~~~~~~~~~~~~(encoding/decoding)& Convolution code ($\frac{1}{3}$ rate) / Viterbi decoding & Sparse encoding  / CS recovery algorithm \\ 
		\hline 
		Transmission & Time/frequency mapping & Spreading in frequency direction\\
		\hline 
		User identification & CRC scrambled with user index & User codebook $\mathbf{C}$ \\ 
		\hline 
		Resource overhead ($L$ repetitions, QPSK) & $L \frac{3b}{2}$ & $Lm$ where $m$ is the size of spreading length\\ 
		\hline 
	\end{tabular} 
\end{table}
\subsection{SVC Decoding}
\subsubsection{Support Identification}
As mentioned, the SVC decoding is done by the identification of the support and any sparse recovery algorithm can be employed for this purpose. In this work, we employ the greedy sparse recovery algorithm in the decoding of the SVC-encoded packet. After pre-multiplying the diagonal matrix constructed by the complex exponential $e^{j \measuredangle h}$, the modified received vector can be expressed as
\begin{eqnarray}
\tilde{\mathbf{y}} &=& \text{diag}\left[ \exp(j \measuredangle h_{11}) \dots \exp(j\measuredangle h_{mm})  \right]{\mathbf{y}} \nonumber \\
&=& \text{diag} \left[\tilde{\mathbf{h}}\right]\mathbf{Cs}+\tilde{\mathbf{v}}, \nonumber \\
&=& \tilde{\mathbf{H}}\mathbf{Cs}+\tilde{\mathbf{v}},
\end{eqnarray}
where $\measuredangle h$ is the angle of $h$,  $\tilde{\mathbf{h}}= [h_{11} e^{j \measuredangle h_{11}} \dots h_{mm} e^{j \measuredangle h_{mm}}]$, $\tilde{\mathbf{H}}= \text{diag}\left[\tilde{\mathbf{h}}\right]$, and $\tilde{\mathbf{v}}=[\tilde{v}_1, \dots, \tilde{v}_m]$ is the modified noise vector where $\tilde{v}_{i}=v_i e^{j \measuredangle h_{ii}}$. Since $\mathbf{s}$ has $K$ nonzero elements, the modified received vector $\tilde{\mathbf{y}}=\tilde{\mathbf{H}}\mathbf{Cs}+\tilde{\mathbf{v}}$ can be expressed as a linear combination of $K$ columns of $\mathbf{\Phi}=\tilde{\mathbf{H}}\mathbf{C}$ perturbed by the noise.
In view of this, the main task of the SVC decoding is to identify the columns in $\mathbf{\Phi}$ participating in the modified received vector. In each iteration, greedy sparse recovery algorithm identifies one column of  $\mathbf{\Phi}$ at a time using a greedy strategy \cite{gomp}. Specifically, a column of $\mathbf{\Phi}$ that is maximally correlated with the (modified) observation $\mathbf{r}^{j-1}$ is chosen. That is, an index of the nonzero column of $\mathbf{\Phi}$ chosen as $j$-th iteration is
\footnote{If ${\Omega}=\left\{ 1, 3\right\}$, then $\mathbf{\Phi}_{\Omega}=[\bm{\phi}_1~\bm{\phi}_3]$.}
\begin{equation}
\label{eq:7}
\omega_j= \arg \underset{l}{\max} ~ |\!\!<\!\bm{\phi}_l, \mathbf{r}^{j-1} \!>\!\!|^2,
\end{equation}
where  $\mathbf{r}^{j-1}=\tilde{\mathbf{y}}-\mathbf{\Phi}_{\Omega^{j-1}_\mathbf{s}}\hat{\mathbf{s}}^{j-1}$ is the modified observation called the residual and $\hat{\mathbf{s}}^{j-1}=\mathbf{\Phi}_{\Omega^{j-1}_\mathbf{s}}^{\dagger}\tilde{\mathbf{y}}$ is the estimate of $\mathbf{s}$ at $(j-1)$-th iteration.\footnote{$\mathbf{\Phi}^{\dagger}= (\mathbf{\Phi}^T \mathbf{\Phi})^{-1}\mathbf{\Phi}^T$ is the pseudo-inverse of $\mathbf{\Phi}$.}

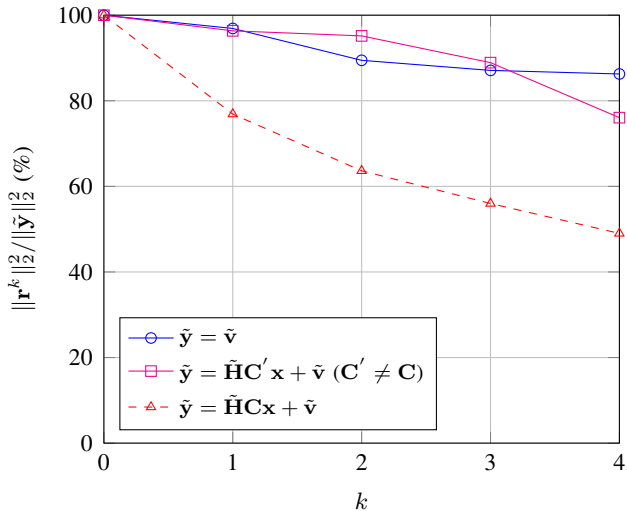
\begin{figure}[t!]
	\centering
	\begin{tikzpicture}
	\begin{axis}[
	xlabel=\footnotesize{$k$},
	ylabel=\footnotesize{$\|\mathbf{r}^k\|^2_2/\|\tilde{\mathbf{y}}\|^2_2$ (\%)},
	xmin=0, xmax=4,
	ymin=0, ymax=100,
	grid={major},
	xtick={0, 1, 2, 3, 4},
	xticklabels={\footnotesize{0}, \footnotesize{1}, \footnotesize{2}, \footnotesize{3},  \footnotesize{4}},
	ytick={0, 20, 40, 60, 80, 100},
	yticklabels={\footnotesize{0}, \footnotesize{20}, \footnotesize{40}, \footnotesize{60}, \footnotesize{80}, \footnotesize{100}},
	legend cell align=left,
	legend style={legend pos=south west},
	scale=1.0
	]
	\addplot[color=blue,mark=o, mark options={solid}] coordinates {
		(0, 100*6.5668/6.5668) (1, 100*6.3632/6.5668) (2, 100*5.8742/6.5668) (3, 100*5.7198/6.5668) (4,100*5.6648/6.5668)
	};
	\addplot[color=magenta, mark=square, mark options={solid}] coordinates {
		(0, 100*8.48/8.48) (1, 100*8.17/8.48) (2, 100*8.07/8.48) (3, 100*7.54/8.48) (4, 100*6.45/8.48)
	};
	\addplot[dashed, color=red,mark=triangle, mark options={solid}] coordinates {
		(0, 100*10.5752/10.5752) (1, 100*8.1313/10.5752) (2, 100*6.7297/10.5752) (3, 100*5.9219/10.5752) (4, 100*5.1829/10.5752)
	};
	\legend{\scriptsize {$\tilde{\mathbf{y}}=\tilde{\mathbf{v}}$}, \scriptsize {$\tilde{\mathbf{y}}=\tilde{\mathbf{H}}\mathbf{C^{'}x}+ \tilde{\mathbf{v}}$ ($\mathbf{C^{'}}\neq\mathbf{C}$)},    \scriptsize{$\tilde{\mathbf{y}}=\tilde{\mathbf{H}}\mathbf{Cx}+\tilde{\mathbf{v}}$}}
	\end{axis}
	\end{tikzpicture}
	
	\caption{Snapshot of the ratio between residual magnitude $\|\mathbf{r}^k\|^2_2$ and $\|\tilde{\mathbf{y}}\|^2_2$ as a function of the number of iterations in the OMP algorithm. Signal-to-noise ratio ($\mathsf{SNR}$) is set to $0$ dB and the sparsity $K$ is set to 4.}
	\label{fig:result21}
\end{figure}
A better way to improve the decoding performance is to use the maximum likelihood (ML) detection. Recalling that the sparsity $K$ is known to both transmitter and receiver, the ML detection problem for the system model in (\ref{eq:7}) is 
\begin{equation}
{\mathbf{s}^*} = \arg \max_{ \| \mathbf{s} \|_0 = K } {P}_r ( \tilde{\mathbf{y}} | \mathbf{s}, \tilde{\mathbf{H}}, \mathbf{C}),
\end{equation}
where $\|\mathbf{s}\|_0$ is the $\ell_0$-norm of $\mathbf{s}$ counting the number of nonzero elements in $\mathbf{s}$.
Since our goal is to find out the support of $\mathbf{s}$, we alternatively have
\begin{equation}
{{\Omega}^*_\mathbf{s}} = \arg \max_{ | {\Omega}_\mathbf{s} | = K } P_r ( \tilde{\mathbf{y}} | {\Omega}_\mathbf{s}, \tilde{\mathbf{H}}, \mathbf{C}),
\end{equation}
where $| {\Omega}_\mathbf{s} |$ is the cardinality of the set $ {\Omega}_\mathbf{s} $.

To find out the ML solution, we need to enumerate all possible combinations of candidate supports with cardinality $K$. Unfortunately, this exhaustive search would not be feasible for most practical scenarios. In this work, we instead use the multipath match pursuit (MMP) algorithm  \cite{mmp}, a recently proposed near-ML sparse recovery algorithm, as a baseline for the SVC decoding. In a nutshell, MMP performs an efficient tree search to find out the near-ML solution to the original sparse vector. Unlike the single-path search algorithm, MMP selects multiple promising indices in each iteration. Specifically, each candidate chosen in an iteration brings forth multiple new child candidates. After finishing $K$ iterations, candidate $\mathbf{s}^{*}$ having the smallest cost function among all candidates is chosen as the final output (i.e., $\mathbf{s}^{*}= \arg \underset{\mathbf{\hat{s}}}{\min} J(\mathbf{\hat{s}})$ where $J(\mathbf{\hat{s}})= \| \tilde{\mathbf{y}} - \mathbf{\Phi}_{{\Omega_{\mathbf{\hat{s}}}}}\mathbf{\hat{s}} \|_2$). Due to the fact that many candidates are redundant and hence counted only once, an actual number of candidates examined in MMP are quite moderate \cite{mmp}.

One clear advantage of MMP, in the perspective of SVC decoding, is that it deteriorates the quality of incorrect candidate yet does not impose any estimation error to the correct one. This is because the quality of incorrect candidates gets worse due to the error propagation while no such behavior occurs to the correct one. In particular, since nonzero values of an original sparse vector $\mathbf{s}$ are known to the receiver,\footnote{Since the goal of SVC decoding is to find out the nonzero positions of a sparse vector, we can pre-define values of the nonzero elements in $\mathbf{s}$ (see Section IV.A).} no estimation error will be introduced in the correct candidate. We note that the computational complexity of the SVC decoding is marginal since the computational complexity of the greedy sparse recovery algorithm is directly proportional to the sparsity $K$.\footnote{In each iteration, greedy sparse recovery algorithm performs three operations: support identification, nonzero element estimation, and residual update. Since the nonzero values are fixed and known in advance, estimation of the nonzero elements is unnecessary.}  Accordingly, the processing latency of SVC decoding can also be made sufficiently small. This is in contrast to the Viterbi or Turbo decoding algorithm in which the computational complexity is proportional to the length of a codeblock \cite{coding}.

{\setstretch{3.0}
	\begin{table}[t!]\footnotesize
		\centering
		\caption{The proposed MMP-based SVC decoding algorithm}
		\begin{tabular}{p{9cm}}
			\hline \hline
			\textbf{Input:} \\ 
			~~Measurement $\tilde{\mathbf{y}}$, sensing matrix $\mathbf{\Phi}=\tilde{\mathbf{H}}\mathbf{C}$, sparsity $K$, \\
			~~number of expansion $L$, max number of search candidate $l_{max}$,\\
			~~stop threshold $\epsilon$, detection threshold $\varepsilon$\\ 
			\textbf{Output:} \\ 
			~~Support set ${\hat{\Omega}}$\\
			\textbf{Initialization:}\\
			~~$l:=0$ (candidate order), $\rho:= \infty$ (minimum magnitude of residual)\\  
			\hline 
			\textbf{While:} $l<l_{max}$ and $\epsilon < \rho$ \textbf{do}\\
			~~$l:=l+1$\\
			~~$\mathbf{r}^0:=\tilde{\mathbf{y}}$\\
			~~$[p_1, ...~,p_K]:= \text{compute\_p}_k(l, L)$~~~~~~~~~~~~~~~~~({\it compute layer order})\\
			~~\textbf{for} $k=1$ \textbf{to} $K$ \textbf{do}~~~~~~~~~~~~~~~~~~~~~~~~~~~~({\it investigate $l$-th candidate})\\
			~~~~$\tilde{\omega}:=$compute\_$\omega(k, L)$~~~~~~~~~~~~~~~~~~~~~~~~({\it choose L best indices})\\
			~~~~${\Omega}^k_l:={\Omega}^{k-1}_l \cup \{ \tilde{\omega}_{p_k}\}$~~~~~~~~~~~~~~~~~~~({\it construct a path in k-th layer})\\
			~~~~$\mathbf{r}^k:=\tilde{\mathbf{y}}-\mathbf{\Phi}_{{\Omega}^k_l}\mathbf{{s}}^k$~~~~~~~~~~~~~~~~~~~~~~~~~~~~~~~~~~~~({\it update residual})\\
			~~~~${\hat{\Omega}}^k:={\Omega}^k_l$~~~~~~~~~~~~~~~~~~~~~~~~~~~~~~~~~~~~~~~~~~({\it {update support set}})\\
			~~\textbf{end for}\\
			~~\textbf{if} $\|\mathbf{r}^K\|_2^2<\rho$ \textbf{then} ~~~~~~~~~~~~~~~~~~~~~~~~~~({\it update the smallest residual})\\
			~~~~$\rho:=\|\mathbf{r}^K\|^2_2$\\
			~~~~~~\textbf{if} $\frac{\|\mathbf{r}^K\|_2^2}{\|\mathbf{y}\|_2^2}>1-\varepsilon$ \textbf{then} ~~~~~~~~~~~~~~~~~~~~~~({\it {false-alarm identification}})\\
			~~~~~~~~${\hat{\Omega}}^*:=\mathbf{0}$\\
			~~~~~~\textbf{end if}\\
			~~~~${\hat{\Omega}}^*:={\hat{\Omega}}^K$\\
			~~\textbf{end if}\\
			\textbf{end while}\\
			\textbf{return} ${\hat{\Omega}}^*$\\
			\hline 
			\textbf{function} $\text{compute\_p}_k (l, L)$ \\
			~~$t:=l-1$\\
			~~\textbf{for} $k=1$ \textbf{to} $K$ \textbf{do}\\
			~~~~$p_k:=\mod(t,L)+1$\\
			~~~~$t:=\text{floor}(t/L)$\\
			~~\textbf{end for}
			\textbf{return} $[p_1,~...~,p_K]$\\
			\textbf{end function}\\
			\hline 
			\textbf{function} $\text{compute\_}\omega(k, L)$ \\
			~\textbf{if} $k=\text{odd}$~\textbf{then} \\
			~~\textbf{return} $\arg \underset{|\pi|=L}{\max} \| (\Re\langle \frac{\bm{\phi}^T}{\|\bm{\phi}\|_2} \mathbf{r}^{k-1}\rangle)_{\pi}\|^2_2$\\
			~\textbf{else}\\
			~~\textbf{return} $\arg \underset{|\pi|=L}{\max} \| (\Im\langle \frac{\bm{\phi}^T}{\|\bm{\phi}\|_2} \mathbf{r}^{k-1}\rangle)_{\pi}\|^2_2$\\
			~\textbf{end if} \\
			\textbf{end function}\\
			\hline \hline
		\end{tabular} 
	\end{table}
}

\subsubsection{Identification of False Alarm}
Overall, there are two kinds of false alarm events causing the decoding failure: 1) support detection when the basestation transmits information to the different user and 2) support detection when there is no transmission at the basestation. In order to prevent these events, we need to examine the residual magnitude in each iteration. Firstly, when a packet for the different user is received, the codebook between two distinct users would be different from each other so that the magnitude of the correlation $\mu_{ij}$ between two codewords, each being chosen from two district codebooks would be small. In this case, clearly, one cannot expect a substantial reduction in the residual magnitude. Secondly, when there is no transmission, the received vector will measure the noise only (i.e., $\tilde{\mathbf{y}}=\tilde{\mathbf{v}}$) and thus some column in $\mathbf{\Phi}$, say $\bm{\phi}_l$, will be added to the residual in each iteration $\mathbf{r}^i = \mathbf{r}^{i-1} - \bm{\phi}_l \mathbf{\hat{s}}_l$ (see Fig. \ref{fig:result21}). Based on these observations, we declare the decoding failure when the residual magnitude is outside of the confidence interval of the pure noise contribution. We will say more about the selection of confidence interval Section V.E.

The proposed MMP-based SVC decoding algorithm is summarized in Table III.

\section{SVC Performance Analysis}
In this section, we analyze the decoding success probability of the SVC technique. As mentioned, decoding of the SVC-encoded packet is successful when all support elements are chosen by the sparse recovery algorithm so that we analyze the probability that the support is identified accurately. In our analysis, we assume that the greedy sparse recovery algorithm is used in the decoding process and analyze the lower bound of the success probability. For analytic simplicity, we initially consider $K\!=\!2$ scenario and then extend to the general case. Without loss of generality, we assume that $p$ and $q$-th elements of $\mathbf{s}$ are nonzero (i.e., $\Omega_\mathbf{s}=\{p, q\}$). Further, by setting the information vector such that $s_p=1$ and $s_q=j$, we can model the QPSK transmission (see Section V.B). 

Following lemmas will be useful in our analysis.
\begin{lemma}
Consider the vector $\mathbf{a}_i$ ($i=1, \cdots, N$) whose element is i.i.d.  standard Gaussian. Then, $\frac{\mathbf{a}_i^T\mathbf{a}_j}{\|\mathbf{a}_i\|_2}$ is standard  Gaussian. That is, $\frac{\mathbf{a}_i^T\mathbf{a}_j}{\|\mathbf{a}_i\|_2} \sim \mathcal{N}(0, 1)$.
\end{lemma}
\begin{IEEEproof}
	See Appendix A.  
\end{IEEEproof}
\begin{lemma}
	Consider the vector $\tilde{\mathbf{h}}=[\tilde{h}_{11}~\tilde{h}_{22}~\cdots ~ \tilde{h}_{mm}]^T$ where $\tilde{h}_{ii} = h_{ii} e^{j \measuredangle h_{ii}}$. The probability density function  (PDF) of the $\|\tilde{\mathbf{h}}\|^2_2$ is Chi-squared distribution with 
	\begin{eqnarray}
	 f_{\|\tilde{\mathbf{h}}\|^2_2}(x)= {\frac  {x^{{m-1}}\exp \left( -x\right)} { \Gamma ({  {m}{}})}},
	\end{eqnarray}
	where $\Gamma (m) =(m-1)! $ is the Gamma function and $\mathrm{E}\left[\|\tilde{\mathbf{h}}\|^2_2\right]=m$.
\end{lemma} 
\begin{IEEEproof}
	From (7), $\|\tilde{\mathbf{h}}\|^2_2$ can be expressed as $\|\tilde{\mathbf{h}}\|^2_2 = \|{\mathbf{h}}\|^2_2 = \sum_{i=1}^{m} |h_{ii}|^2=\sum_{i=1}^{m}(\Re(h_{ii})^2 + \Im(h_{ii})^2)$ where $\Re (c)$ and  $\Im (c)$  are the real and imaginary part of $c$, respectively. Since $\Re(h_{ii})$, $\Im(h_{ii})$ $\sim \mathcal{N}(0, \frac{\sigma_v^2}{2})$, we can show after some manipulations that $2\|\tilde{\mathbf{h}}\|^2_2$ follows Chi-squared distribution with  $2m$ DoF \cite{book}. That is, 
	\begin{eqnarray}
	f_{2\|\tilde{\mathbf{h}}\|^2_2}(x)= {\frac  {x^{{m-1}}\exp \left( -\frac{x}{2}\right)} {2^m \Gamma ({  {m}{}})}}.
	\end{eqnarray}	
	Since $f_Z(z)=2f_{2Z}(2z)$, we have
		\begin{eqnarray}
	f_{\|\tilde{\mathbf{h}}\|^2_2}(x)= {\frac  {x^{{m-1}}\exp \left( -x\right)} {\Gamma ({  {m}{}})}}.
	\end{eqnarray}	
\end{IEEEproof}
	
Let $\mathcal{S}^j$ be the success probability that the support element is chosen in the $j$-th iteration. Since $K=2$ and thus the required number of iterations to decode the information vector is two, the probability that the SVC packet is successfully decoded can be expressed as 
	\begin{eqnarray}
	\label{eq:a2}
	P_{succ} &=& \mathrm{P}( {\Omega}^{*}_{\mathbf{s}} = \Omega_\mathbf{s}) \nonumber \\
	&=& \mathrm{P}\left(\mathcal{S}^1, \mathcal{S}^2 \right) \nonumber \\
	&=& \mathrm{P}\left( \mathcal{S}^2 |\mathcal{S}^1\right) \mathrm{P}\left( \mathcal{S}^1 \right).
	\end{eqnarray}
	
Our main result in this section is as follows.

\addtocounter{theorem}{-2}
\begin{theorem}
	The probability that the SVC-encoded packet is decoded successfully satisfies 
	\begin{eqnarray}
	P_{succ} &\geq&  \left( 1-\left(1+\frac{(1-\mu^{*})^2}{\sigma_v^2}\right)^{-m} -\left(1+\frac{1}{\sigma_v^2}\right)^{-m} \right)^{2N}_{\textstyle ~~,}
	\end{eqnarray}
where $m$ is the number of measurements (resources), $N$ is the size of sparse vectors, $\sigma_{v}^2$ is the  noise variance, and $\mu^{*} = \underset{{i\neq j}}{\max}~| \mu_{ij}| $ is the maximum absolute value of correlation between two distinct columns of $\mathbf{\Phi}$.
\end{theorem}
When $m$ is sufficiently large, we approximately have 
\begin{eqnarray}
\label{eq:a51}
{P_{succ}}  &\gtrsim&  \left(1-\left(1+\frac{(1-\mu^{*})^2}{\sigma_v^2}\right)^{-m} \right)^{2N}_{\textstyle ~~.}
\end{eqnarray}
\addtocounter{theorem}{+1}
 Also, since the block error rate is $\text{BLER}_{svc} = 1- P_{succ}$, the upper bound of BLER is   
\begin{eqnarray}
\label{eq:a211}
\text{BLER}_{svc} 
&\lesssim& 1-  \left(1-\left(1+\frac{(1-\mu^{*})^2}{\sigma_v^2}\right)^{-m} \right)^{2N}_{\textstyle ~~.}
\end{eqnarray} 
\begin{figure}[t!]
	\centering
		{
		\begin{tikzpicture}
		\begin{semilogyaxis}[
		xlabel=\footnotesize{SNR (dB)},
		ylabel=\footnotesize{BLER},
		xmin=-0, xmax=8,
		ymin=10e-6, ymax=1,
		grid={major},
		xtick={0 , 2, 4, 6,8,10},
		xticklabels={ \footnotesize{0},  \footnotesize{2} , \footnotesize{4}, \footnotesize{6}, \footnotesize{8}},
		ytick={10e-6, 10e-5, 10e-4, 10e-3, 10e-2,10e-1, 1},
		yticklabels={\footnotesize{$10^{-5}$}, \footnotesize{$10^{-4}$}, \footnotesize{$10^{-3}$}, \footnotesize{$10^{-2}$}, \footnotesize{$10^{-1}$}, \footnotesize{$1$}},
		legend cell align=left,
		legend style={legend pos=north east},
		scale=0.85
		]
		\addplot[dashed, color=red,mark=triangle*, mark options={solid}] coordinates {
			(-4+2,	1) 
			(-3+2,	1) 
			(-2+2, 	1)
			(-1+2,	0.999)
			(-0.5+2, 0.8187)
			(0+2,	0.5176)	
			(0.5+2, 0.1996)
			(1+2, 0.01715)
			(1.5+2., 0.0011)
			(3+2., 6.6667e-09)
			(4+2., 1e-13)
		};
		\addplot[color=blue,mark=star, mark options={solid}] coordinates {
			(-5+0.7+5,	1)
			(-4.5+0.7+5,	1) (-4.4+0.7+5,	1) (-4.2+0.7+5,	1)
			(-4+0.7+5,	0.9999929) (-3.8+0.7+5, 0.998042) (-3.6+0.7+5, 0.954182)
			(-3.5+0.7+5,	0.879278556) (-3.4+0.7+5,	0.75934) (-3.2+0.7+5,	0.457107)
			(-3+0.7+5,	0.214489039) (-2.8+0.7+5,	0.083546)
			(-2.5+0.7+5,	0.015596352)
			(-2+0.7+5,	0.000504442)
			(-1.5+0.7+5,	6.64816E-06)
			(-1+0.7+5,	2.85529E-08)
			(-0.5+0.7+5,	2.96723E-11)
			(0+0.7+5,	0)
		};
		\legend{ \scriptsize {Simulation}, \scriptsize {Upper bound in (\ref{eq:a211})}}
		\end{semilogyaxis}
		\end{tikzpicture}
	}
	{
		\begin{tikzpicture}
		\begin{semilogyaxis}[
		xlabel=\footnotesize{SNR (dB)},
		ylabel=\footnotesize{BLER},
		xmin=-4, xmax=6,
		ymin=10e-6, ymax=1,
		grid={major},
		xtick={-4, -2,   0  , 2, 4, 6,8,10},
		xticklabels={\footnotesize{-4}, \footnotesize{-2}, \footnotesize{0},  \footnotesize{2} , \footnotesize{4}, \footnotesize{6}},
		ytick={10e-6, 10e-5, 10e-4, 10e-3, 10e-2,10e-1, 1},
		yticklabels={\footnotesize{$10^{-5}$}, \footnotesize{$10^{-4}$}, \footnotesize{$10^{-3}$}, \footnotesize{$10^{-2}$}, \footnotesize{$10^{-1}$}, \footnotesize{$1$}},
		legend cell align=left,
		legend style={legend pos=south east},
		scale=0.85
		]
		\addplot[color=blue,mark=star, mark options={solid}] coordinates {
			(-5,	1)
			(-4.5,	1) (-4.4,	1) (-4.2,	1)
			(-4,	0.9999929) (-3.8, 0.998042) (-3.6, 0.954182)
			(-3.5,	0.879278556) (-3.4,	0.75934) (-3.2,	0.457107)
			(-3,	0.214489039) (-2.8,	0.083546)
			(-2.5,	0.015596352)
			(-2,	0.000504442)
			(-1.5,	6.64816E-06)
			(-1,	2.85529E-08)
			(-0.5,	2.96723E-11)
			(0,	0)
		};
		\addplot[color=red,mark=triangle, mark options={solid}] coordinates {
			(-4,	1) (-3.5,	1)
			(-3,	0.999979107) (-2.8, 0.996361) (-2.6, 0.936004)
			(-2.5,	0.846641322) (-2.4, 0.715219) (-2.2, 0.412664)
			(-2,	0.187422312)
			(-1.5,	0.012910607)
			(-1,	0.000396976)
			(-0.5,	4.91682E-06)
			(0,	1.95305E-08)
			(0.5,	1.84173E-11)
			(1,	0)
		};
		\addplot[color=black,mark=x, mark options={solid}] coordinates {
			(-3,	1) (-2.5,	1) (-2,	1) (-1.8, 0.999996) (-1.6, 0.998685)
			(-1.5,	0.99105079) (-1.4, 0.9631) (-1.2, 0.784411) 
			(-1,	0.484573084) (-0.8, 0.232033)
			(-0.5,	0.054506425)
			(0,	0.002497279)
			(0.5,	4.98605E-05)
			(1,	3.60819E-07)
			(1.5,	7.28505E-10)
			(2,	0)	
		};
		\addplot[color=magenta,mark=o, mark options={solid}] coordinates {
			(-1.5,	1)  (-1,	1)
			(-0.5,	1)
			(0,	0.999985278) (0.2, 0.997022) (0.4, 0.942537)
			(0.5,	0.858001607) (0.6, 0.730199) (0.8, 0.427254)
			(1,	0.196135238) (1.2, 0.07516) (1.4, 0.025006)
			(1.5,	0.01375459)
			(2,	0.000430142)
			(2.5,	5.43956E-06)
			(3,	2.21795E-08)
			(3.5,	2.14868E-11)
			(4,	0)
		};
		\addplot[color=green,mark=square, mark options={solid}] coordinates {
			( 0.5,	1)( 1,	1)( 1.5,	1)( 2,	1)( 2.5,	1)
			( 3	,0.999989726) (3.2, 0.997577195) (3.4, 0.948591467)
			( 3.5,	0.868884416) (3.6, 0.744915841) (3.8, 0.442073454)
			( 4,0.205156198) (4.2, 0.079256773) (4.4, 0.026547273)
			(4.5,	0.014648948) (4.6, 0.00783726)
			( 5,	0.000465901)
			(5.5,	6.015E-06)
			( 6,	2.51723E-08)
		};
		\legend{\scriptsize {$\mu^{*}=0.0$}, \scriptsize {$\mu^{*}=0.2$}, \scriptsize {$\mu^{*}=0.4$},  \scriptsize {$\mu^{*}=0.6$},  \scriptsize {$\mu^{*}=0.8$}}
		\end{semilogyaxis}
		\end{tikzpicture}
		
			{\footnotesize ~~~(a) the exact BLER and bound ($m=42$, $N=96$)~~~~~~~~~~~~ (b) BLER with $\mu^{*}$ ($m=42$, $N=96$) }
			
	}
	~\\
		{
			\begin{tikzpicture}
			\begin{semilogyaxis}[
			xlabel=\footnotesize{SNR (dB)},
			ylabel=\footnotesize{BLER},
			xmin=-3.8, xmax=0.2,
			ymin=10e-6, ymax=1,
			grid={major},
			xtick={-3.8, -2.8,   -1.8  , -0.8, 0.2},
			xticklabels={\footnotesize{1}, \footnotesize{2}, \footnotesize{3},  \footnotesize{4} , \footnotesize{5}},
			ytick={10e-6, 10e-5, 10e-4, 10e-3, 10e-2,10e-1, 1},
			yticklabels={\footnotesize{$10^{-5}$}, \footnotesize{$10^{-4}$}, \footnotesize{$10^{-3}$}, \footnotesize{$10^{-2}$}, \footnotesize{$10^{-1}$}, \footnotesize{$1$}},
			legend cell align=left,
			legend style={legend pos=south west},
			scale=0.85
			]
			\addplot[color=blue,mark=star, mark options={solid}] coordinates {
				(-3.4,	1)
				(-3.2,	1)
				(-3,	1)
				(-2.8,	0.999996748)
				(-2.6,	0.997940134)
				(-2.4,	0.940755296)
				(-2.2,	0.698008943)
				(-2,	0.373103263)
				(-1.8,	0.153214451)
				(-1.6,	0.052201914)
				(-1.4,	0.015376185)
				(-1.2,	0.003965257)
				(-1,	0.000892974)
				(-0.8,	0.000173927)
				(-0.6,	2.89207E-05)
				(-0.4,	4.04437E-06)
				(-0.2,	4.67801E-07)
				(0,	4.39436E-08)
				(0.2,	3.28518E-09)
				(0.4,	1.91079E-10)
				(0.6,	9.20863E-12)
				(0.8,	0)
				
			};
			\addplot[color=red,mark=triangle, mark options={solid}] coordinates {
				(-4,	1) (-3.5,	1)
				(-3,	0.999979107) (-2.8, 0.996361) (-2.6, 0.936004)
				(-2.5,	0.846641322) (-2.4, 0.715219) (-2.2, 0.412664)
				(-2,	0.187422312)
				(-1.5,	0.012910607)
				(-1,	0.000396976)
				(-0.5,	4.91682E-06)
				(0,	1.95305E-08)
				(0.5,	1.84173E-11)
				(1,	0)
				
			};
			\addplot[color=black,mark=x, mark options={solid}] coordinates {
				(-3.4,	0.999999673)
				(-3.2,	0.999829978)
				(-3,	0.99168235)
				(-2.8,	0.917589015)
				(-2.6,	0.70528633)
				(-2.4,	0.427783466)
				(-2.2,	0.210626066)
				(-2,	0.088115251)
				(-1.8,	0.032317193)
				(-1.6,	0.010534497)
				(-1.4,	0.003056185)
				(-1.2,	0.00078451)
				(-1,	0.000176453)
				(-0.8,	3.43583E-05)
				(-0.6,	5.7128E-06)
				(-0.4,	7.98888E-07)
				(-0.2,	9.24051E-08)
				(0,	8.68022E-09)
				(0.2,	6.48924E-10)
				(0.4,	3.7744E-11)
				(0.6,	1.81899E-12)
				(0.8,	0)
			};
			\addplot[color=green,mark=o, mark options={solid}] coordinates {
				(-3.4,	0.999775066)
				(-3.2,	0.992420168)
				(-3,	0.932391645)
				(-2.8,	0.754392502)
				(-2.6,	0.497035123)
				(-2.4,	0.269487375)
				(-2.2,	0.124569719)
				(-2,	0.050562845)
				(-1.8,	0.018308962)
				(-1.6,	0.005939379)
				(-1.4,	0.001720255)
				(-1.2,	0.000441362)
				(-1,	9.92587E-05)
				(-0.8,	1.93267E-05)
				(-0.6,	3.21345E-06)
				(-0.4,	4.49375E-07)
				(-0.2,	5.19779E-08)
				(0,	4.88262E-09)
				(0.2,	3.6502E-10)
				(0.4,	2.1231E-11)
				(0.6,	1.02318E-12)
			};
			\legend{\scriptsize {$N=144$},  \scriptsize {$N=96$}, \scriptsize {$N=64$},  \scriptsize {$N=48$}}
			\end{semilogyaxis}
			\end{tikzpicture}
		}
		{
			\begin{tikzpicture}
			\begin{semilogyaxis}[
			xlabel=\footnotesize{SNR (dB)},
			ylabel=\footnotesize{BLER},
			xmin=-4.8, xmax=3.2,
			ymin=10e-6, ymax=1,
			grid={major},
			xtick={-4.8, -2.8,   -0.8  , 1.2, 3.2},
			xticklabels={\footnotesize{0}, \footnotesize{2}, \footnotesize{4},  \footnotesize{6},  \footnotesize{8} },
			ytick={10e-6, 10e-5, 10e-4, 10e-3, 10e-2,10e-1, 1},
			yticklabels={\footnotesize{$10^{-5}$}, \footnotesize{$10^{-4}$}, \footnotesize{$10^{-3}$}, \footnotesize{$10^{-2}$}, \footnotesize{$10^{-1}$}, \footnotesize{$1$}},
			legend cell align=left,
			legend style={legend pos=south west},
			scale=0.85
			]
			\addplot[color=green,mark=square, mark options={solid}] coordinates {
				(-6,	1) (-5.8,	1) (-5.6,	1) (-5.4,	1) (-5.2,	1) (-5,	1) (-4.8,	1) (-4.6,	1)  (-4.4,	1) (-4.2,	1)(-4,	0.99999981)
				(-3.8,	0.999763751)
				(-3.6,	0.985686607)
				(-3.4,	0.867790756)
				(-3.2,	0.592452468)
				(-3,	0.308000661)
				(-2.8,	0.129401727)
				(-2.6,	0.046356888)
				(-2.4,	0.014554842)
				(-2.2,	0.004035473)
				(-2,	0.000984401)
				(-1.8,	0.000209303)
				(-1.6,	3.83143E-05)
				(-1.4,	5.95343E-06)
			};
			\addplot[color=red,mark=triangle, mark options={solid}] coordinates {
				(-4,	1) (-3.5,	1)
				(-3,	0.999979107) (-2.8, 0.996361) (-2.6, 0.936004)
				(-2.5,	0.846641322) (-2.4, 0.715219) (-2.2, 0.412664)
				(-2,	0.187422312)
				(-1.5,	0.012910607)
				(-1,	0.000396976)
				(-0.5,	4.91682E-06)
				(0,	1.95305E-08)
				(0.5,	1.84173E-11)
				(1,	0)
				
			};
			%
			\addplot[color=magenta,mark=o, mark options={solid}] coordinates {
				(-2.4,	1)
				(-2.2,	0.999998962)
				(-2,	0.999358891)
				(-1.8,	0.975111435)
				(-1.6,	0.823823744)
				(-1.4,	0.531844842)
				(-1.2,	0.263838681)
				(-1,	0.10708316)
				(-0.8,	0.037333056)
				(-0.6,	0.01143862)
				(-0.4,	0.003094715)
				(-0.2,	0.00073554)
				(0,	0.000152038)
				(0.2,	2.69856E-05)
				(0.4,	4.05368E-06)
			};
			\addplot[color=blue,mark=star, mark options={solid}] coordinates {
				(-1.8,	1)(-1.6,	1)(-1.4,	1)(-1.2,	1)(-1,	1)(-0.8,	1) (-0.6,	1)	(-0.8,	0.999999993)
				(-0.6,	0.999966456)
				(-0.4,	0.995233695)
				(-0.2,	0.92605602)
				(0,	0.693897941)
				(0.2,	0.392689233)
				(0.4,	0.17575409)
				(0.6,	0.066081363)
				(0.8,	0.021636794)
				(1,	0.006250588)
				(1.2,	0.00159195)
				(1.4,	0.000354639)
				(1.6,	6.83116E-05)
				(1.8,	1.12237E-05)
				(2,	1.54915E-06	)
			};
			\legend{\scriptsize {$m=63$}, \scriptsize {$m=42$}, \scriptsize {$m=28$}, \scriptsize {$m=14$} }
			\end{semilogyaxis}
			\end{tikzpicture}

			{\footnotesize ~~~~~~~~~~~~(c) BLER with $N$ ($m=42$, $\mu^{*}=0.7$) ~~~~~~~~~~~~~~~~~~~~~~~ (d)  BLER with $m$ ($N=96$, $\mu^{*}=0.7$)}
		}
	\caption{BLER performance of SVC-encoded packet from (\ref{eq:a211})}
	\label{fig:bler}
\end{figure}
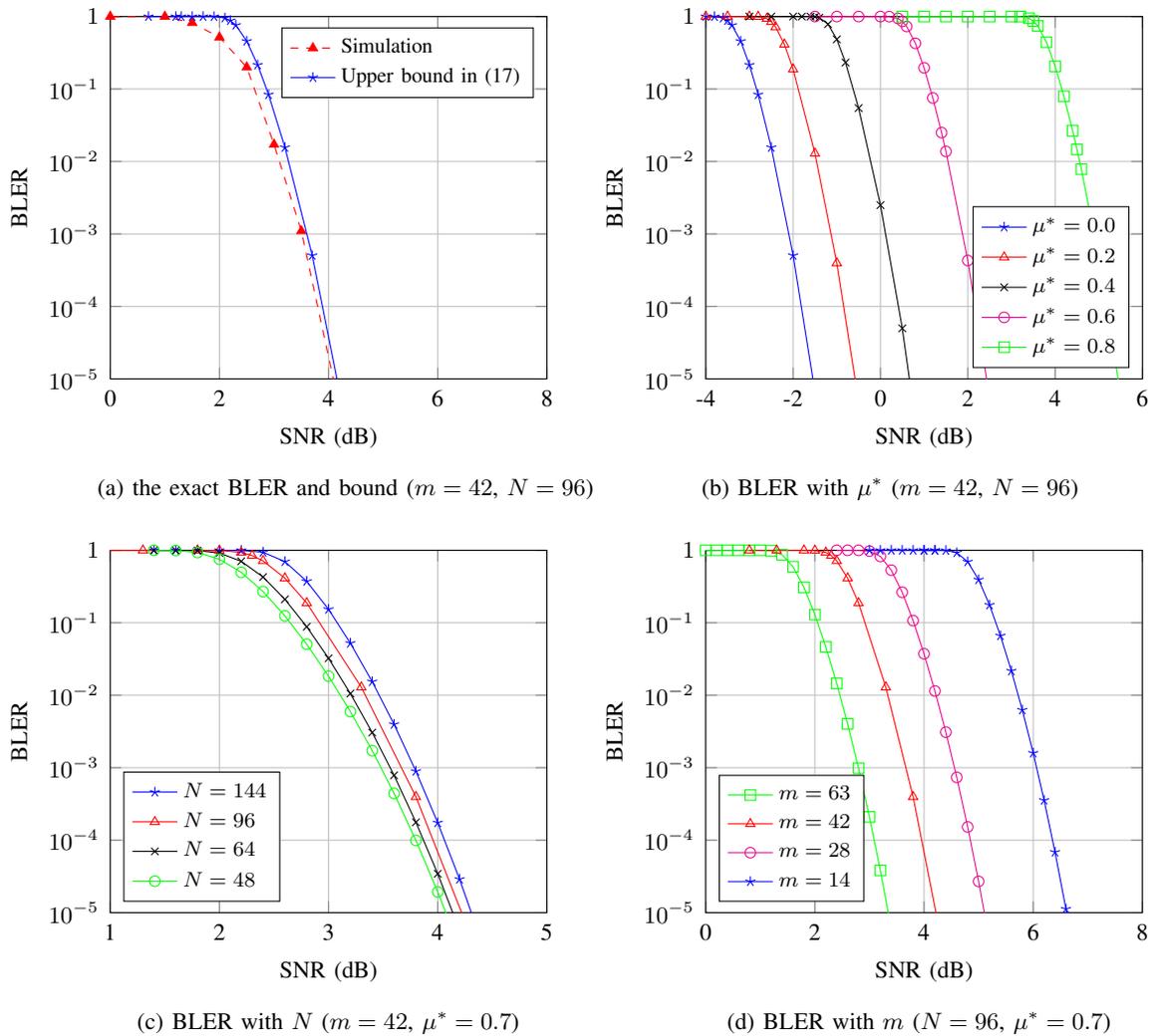

In Fig. \ref{fig:bler}, we plot the BLER performance of SVC as a function of SNR. To judge the effectiveness of Theorem 1, we perform the empirical simulation for $m=42$, $N=96$. From the empirical evaluations, we obtain that $\mu^{*} \approx 0.7$. When we apply this value to the upper bound in (\ref{eq:a211}), we could observe that the obtained bound is tight across the board. To better understand the performance of SVC, we plot the BLER as a function of $\mu^{*}, N,$ and $m$ in Fig. \ref{fig:bler}(b), \ref{fig:bler}(c), and \ref{fig:bler}(d). First, when the maximum correlation $\mu^{*}$ decreases, we see that the BLER gain increase sharply as shown in Fig. 4(b). For example, if $\mu^{*}$ is reduced from 0.4 to 0.2, we can achieve 1.5 dB gain at the target reliability point (BLER =  $10^{\text{-}5}$). Next, we test the BLER performance for various sparse vector dimensions in Fig. \ref{fig:bler}(c). Although the BLER performance degrades with $N$, we see that the degradation is fairly graceful. Whereas, as shown in Fig. \ref{fig:bler}(d), the BLER performance is quite sensitive to the number of measurements. 

As a first step to prove Theorem 1, we analyze the success probability $\mathrm{P}\left( \mathcal{S}^1 \right)$ for the first iteration.
\begin{lemma}
 Consider the received signal ${\tilde{\mathbf{y}}}= \gamma\mathbf{\Phi s} + \tilde{\mathbf{v}}$ where $\gamma = {\frac{\sqrt{\mathsf{SNR}}}{\alpha}}$, $\mathbf{\Phi}=[\bm{\phi}_1~\bm{\phi}_2~\cdots~\bm{\phi}_N]$, and {$\bm{\phi}_i= [\tilde{h}_{11}{c}_{1i}~\tilde{h}_{22}{c}_{2i}~\cdots~\tilde{h}_{mm}{c}_{mi}]^T$.} The probability that the support element is chosen in the first iteration satisfies 
 \begin{eqnarray}
 \mathrm{P}( \mathcal{S}^1) &\geq& \left( 1-\left(1+\frac{(1-\mu^{*})^2}{\sigma_v^2}\right)^{-m} -\left(1+\frac{1}{\sigma_v^2}\right)^{-m} \right)^{N-1}_{\textstyle ~~~~.} 
 \end{eqnarray}
 
\end{lemma}
\begin{IEEEproof}
		As shown in Table II, $N$ decision statistics $\frac{\bm{\phi}_l^T}{\|\bm{\phi}_l\|_2}\mathbf{r}^{k-1}$ ($l=1, \cdots,N$) are computed in each iteration. For analytic simplicity, we take the real part of the decision statistic in the first iteration and the imaginary part in the second iteration.\footnote{This choice is suboptimal but simplifies the analysis.} 
	
	In order to identify the support element in the first iteration, we should have $\left|\Re \langle{\frac{ \bm{\phi}_p}{\| \bm{\phi}_p \|_2}, \mathbf{r}^0} \rangle\right| \geq \underset{i}{\max} \left| \Re \langle  \frac{ \bm{\phi}_i}{\| \bm{\phi}_i \|_2}, {\mathbf{r}^0}  \rangle \right|$ and thus the success probability for a given channel realization $\mathbf{h}$ is
	\begin{eqnarray}
	\label{eq:b1}
	\mathrm{P}( \mathcal{S}^1 | \mathbf{h}) &=&\mathrm{P}\left( \left|\Re \langle  \frac{ \bm{\phi}_p}{\| \bm{\phi}_p \|_2}, {\mathbf{r}^0}\rangle\right| \geq \max_{i} \left| \Re \langle \frac{ \bm{\phi}_i}{\| \bm{\phi}_i \|_2}, {\mathbf{r}^0}  \rangle \right| \right) \nonumber \\
	&=&  \prod_{\mathclap{i=1, i\neq p}}^N \mathrm{P}\left( \left|\Re \langle \frac{ \bm{\phi}_p}{\| \bm{\phi}_p \|_2}, {\mathbf{r}^0} \rangle\right| \geq  \left|\Re \langle \frac{ \bm{\phi}_i}{\| \bm{\phi}_i \|_2}, {\mathbf{r}^0} \rangle\right|\right),
	\end{eqnarray}
	where $\langle \mathbf{a}, \mathbf{b}  \rangle$ is the inner product between two vector $\mathbf{a}$ and $\mathbf{b}$.
	First, noting that $s_p=1$ and $s_q=j$, we have 
	\begin{eqnarray}
	\label{eq:a6}
	\langle {\frac{\bm{\phi}_p}{\|\bm{\phi}_p\|_2}, \mathbf{r}^0}  \rangle &=& \langle \frac{\bm{\phi}_p}{\|\bm{\phi}_p\|_2} , \bm{\phi}_p s_p+  \bm{\phi}_q s_q +\tilde{\mathbf{v}}\rangle \nonumber \\
	&=& \|\tilde{\mathbf{h}}\|_2  + j  \|\tilde{\mathbf{h}}\|_2 \mu_{qp} + \frac{\bm{\phi}_l^T}{\|\bm{\phi}_l\|_2}\tilde{\mathbf{v}},
	\end{eqnarray}
	where the equality follows from (see Appendix B)
	\begin{eqnarray}
	\label{eq:a66}
	\langle \frac{\bm{\phi}_k}{\|\bm{\phi}_k\|_2} , \bm{\phi}_l \rangle  &=& \begin{cases} \|\tilde{\mathbf{h}}\|_2 & \text{for  $k= l$} \\   \|\tilde{\mathbf{h}}\|_2 \mu_{kl} & \text{for $k \neq l$}  \end{cases}_{\textstyle .}
	\end{eqnarray}
	Let $z_p=\Re \left(\frac{\bm{\phi}_p^T}{\|\bm{\phi}_p\|_2}\tilde{\mathbf{v}} \right)$, then 
	\begin{eqnarray}
	\label{eq:a61}
	\Re \langle {\frac{\bm{\phi}_p}{\|\bm{\phi}_p\|_2} , \mathbf{r}^0} \rangle &=& \|\tilde{\mathbf{h}}\|_2 + z_p. 	
	\end{eqnarray}
	In a similar way, we have
	\begin{eqnarray}
	\label{eq:a611}
	\Re \langle \frac{\bm{\phi}_i}{\|\bm{\phi}_i\|_2}, {\mathbf{r}^0} \rangle &=& \|\tilde{\mathbf{h}}\|_2\mu_{ip} + z_i,
	\end{eqnarray}
	and hence
	\begin{eqnarray}
	\label{eq:a7}
	\mathrm{P}\left(  \left| \Re\langle  \frac{\bm{\phi}_p}{\|\bm{\phi}_p\|_2}, {\mathbf{r}^0}\rangle\right| \geq  \left|\Re \langle \frac{\bm{\phi}_i}{\|\bm{\phi}_i\|_2}, {\mathbf{r}^0}  \rangle\right|\right)  
	&=& \mathrm{P}\left(\left| \|\tilde{\mathbf{h}}\|_2 + z_p \right| \geq  \left| \|\tilde{\mathbf{h}}\|_2 \mu_{il} + z_i  \right| \right) \nonumber\\
	&\overset{(a)}{=}& \mathrm{P}\left( \|\tilde{\mathbf{h}}\|_2 + z_p   > \left| \|\tilde{\mathbf{h}}\|_2 \mu_{ip}   + z_i   \right|   \right)\mathrm{P} \left( \|\tilde{\mathbf{h}}\|_2 + z_p >0\right) \nonumber\\ && + \mathrm{P}\left( - \|\tilde{\mathbf{h}}\|_2 - z_p   >  \left| \|\tilde{\mathbf{h}}\|_2\mu_{ip}  + z_i  \right| \right)\mathrm{P} \left(\|\tilde{\mathbf{h}}\|_2 + z_p <0\right) \nonumber \\
	&\geq& \mathrm{P}\left( \|\tilde{\mathbf{h}}\|_2 +  z_p >   \|\tilde{\mathbf{h}}\|_2  |\mu_{ip}|  +| z_i | \right)\mathrm{P}\left(\|\tilde{\mathbf{h}}\|_2 +z_p> 0\right) \nonumber  \\
	&\geq& \mathrm{P}\left( \|\tilde{\mathbf{h}}\|_2 +  z_p >   \mu^{*} \|\tilde{\mathbf{h}}\|_2    + |z_i|  \right)\mathrm{P}\left(\|\tilde{\mathbf{h}}\|_2 +z_p> 0\right), \nonumber \\
	\end{eqnarray}
	where $(a)$ follows from 
	\begin{eqnarray}
	\label{eq:a8}
	\mathrm{P}\left(\left| A \right| \geq  \left|  B \right| \right) &=& \mathrm{P}\left(A  >   |B|  \right)\mathrm{P} \left(A>0\right) +\mathrm{P}\left( - A >  \left|B\right| \right)\mathrm{P} \left(A<0\right).
	\end{eqnarray}
	Since $z_i \sim \mathcal{N}(0, \frac{\mathbf{\sigma}_v^2}{2})$ from Lemma 1, the second term in (\ref{eq:a7}) is lower bounded as 
	\begin{eqnarray}
	\label{eq:a9}
	\mathrm{P} \left(\|\tilde{\mathbf{h}}\|_2 +z_p >  0\right) &=& \mathrm{P} \left(z_p >  -\|\tilde{\mathbf{h}}\|_2 \right) \nonumber \\
	&=& 1 - Q\left(-\frac{\|\tilde{\mathbf{h}}\|_2}{\frac{\sigma_{{v}}}{\sqrt{{2}}} }\right) \nonumber \\
	&\geq&  1- \exp \left( -\frac{\|\tilde{\mathbf{h}}\|_2^2}{\sigma_{{v}}^2} \right) ,
	\end{eqnarray}
	where the last inequality follows from $Q(x) \leq  \exp\left({-\frac{x^2}{2}}\right)$.  In a similar way, the first term in (\ref{eq:a7}) is lower bounded as
	\begin{eqnarray}
	\label{eq:a10}
	{\mathrm{P}\left(\|\tilde{\mathbf{h}}\|_2 +  z_p > \mu^{*} \|\tilde{\mathbf{h}}\|_2  + \left| z_i \right| \right) }   &=& 1- \mathrm{P}\left( \left| z_i \right|  -  z_p \geq (1- \mu^{*}) \|\tilde{\mathbf{h}}\|_2   \right)     \nonumber\\
	&=& 1- \mathrm{P}\left(  z_i  -  z_p \geq (1- \mu^{*}) \|\tilde{\mathbf{h}}\|_2   \right)  \mathrm{P}\left(  z_i  >0  \right) \nonumber \\ && -\mathrm{P}\left(  -z_i  -  z_p \geq (1- \mu^{*}) \|\tilde{\mathbf{h}}\|_2   \right)  \mathrm{P}\left(  z_i  <0  \right)      \nonumber\\
	&\overset{(a)}{=}& 1- 2\mathrm{P}\left(  z_i  -  z_p \geq (1- \mu^{*}) \|\tilde{\mathbf{h}}\|_2   \right)  \mathrm{P}\left(  z_i  >0  \right) \nonumber \\ 
	&\overset{(b)}{\geq}& 1-   Q\left(-\frac{\|\tilde{\mathbf{h}}\|_2(1-\mu^{*})}{{\sigma_{{v}}}{}}\right) \nonumber \\
	&{\geq}& 1- \exp \left(-{\frac{\|\tilde{\mathbf{h}}\|_2^2 (1-\mu^{*})^2}{ 2{\sigma_{{v}}}^2}} \right), 
	\end{eqnarray}
	where $(a)$ is because $-z_i\sim \mathcal{N}(0, \frac{\sigma_v^2}{2})$ and $(b)$ is because $z_i-z_p\sim \mathcal{N}(0, \sigma_v^2)$. By plugging  (\ref{eq:a9}) and (\ref{eq:a10}) into (\ref{eq:a7}), we have
	\begin{eqnarray}
	\label{eq:a11}
	\mathrm{P}\left(  \left| \Re\langle {\frac{\bm{\phi}_p}{\|\bm{\phi}_p\|_2} ,\mathbf{r}^0}  \rangle\right|  \geq  \left|\Re \langle \frac{\bm{\phi}_i}{\|\bm{\phi}_i\|_2} , {\mathbf{r}^0} \rangle\right| \right) \!\!&\geq&\!\!  \left(1- \exp \left(-{\frac{\|\tilde{\mathbf{h}}\|_2^2(1-\mu^{*})^2}{ 2{\sigma_{{v}}}^2}}\right) \right) \left( 1- \exp \left(- \frac{\|\tilde{\mathbf{h}}\|_2^2}{\sigma_{{v}}^2} \right)\right) \nonumber \\
	\!\!&\geq&\!\!  1- \exp \left( { -\frac{ \|\tilde{\mathbf{h}}\|_2^2(1-\mu^{*})^2}{2 {\sigma_{{v}}}^2}} \right)- \exp \left( -\frac{\|\tilde{\mathbf{h}}\|_2^2}{\sigma_{{v}}^2} \right). \nonumber \!\!\! \\ \!\!\!
	\end{eqnarray}
	
	Note that $\mathrm{P}(\mathcal{S}^1 | \mathbf{h})$ in (\ref{eq:b1}) is the success probability in the first iteration for a given channel realization $\mathbf{h}$. In order to obtain the unconditional probability, we need to take expectation with respect to the channel $\mathbf{h}$.  That is,
	\begin{eqnarray}
	\mathrm{P}({\mathcal{S}}^1) \!\!\!&=&\!\!\!\!  \int \mathrm{P}(\mathcal{S}^1| \mathbf{h} ) f_{\mathbf{h}}(x) dx = \mathrm{E}_{\mathbf{h}} \left[ \mathrm{P}(\mathcal{S}^1 | \mathbf{h}) \right].
	\end{eqnarray}
	Thus, 
	\begin{eqnarray}
	\label{eq:a111}
	\mathrm{P}({\mathcal{S}}^1) &=& {\mathrm{E}_{\mathbf{h}} \left[~ \prod_{\mathclap{i=1, i\neq p}}^N \mathrm{P}\left(  \left| \Re\langle \frac{\bm{\phi}_p}{\|\bm{\phi}_p\|_2}, {\mathbf{r}^0}  \rangle\right|  \geq   \left|\Re \langle \frac{\bm{\phi}_i}{\|\bm{\phi}_i\|_2}, {\mathbf{r}^0} \rangle\right| \right)   \mathrel{\stretchto{\mid}{3ex}} \mathbf{h}  \right]} \nonumber \\
	&=&  \prod_{\mathclap{i=1, i\neq p}}^N  \mathrm{E}_{\mathbf{h}}  \left[ \mathrm{P}\left(  \left| \Re\langle \frac{\bm{\phi}_p}{\|\bm{\phi}_p\|_2}, {\mathbf{r}^0}  \rangle\right|  \geq   \left|\Re \langle \frac{\bm{\phi}_i}{\|\bm{\phi}_i\|_2}, {\mathbf{r}^0} \rangle\right| \right)   \mathrel{\stretchto{\mid}{3ex}} \mathbf{h}  \right] \nonumber \\
	&\geq&   \prod_{\mathclap{i=1, i\neq p}}^N   \mathrm{E}_{{\mathbf{h}}} \left[ 1- \exp \left( { -\frac{ \|\tilde{\mathbf{h}}\|_2^2(1-\mu^{*})^2}{2 {\sigma_{{v}}}^2}} \right)- \exp \left( -\frac{\|\tilde{\mathbf{h}}\|_2^2}{\sigma_{{v}}^2} \right) \mathrel{\stretchto{\mid}{4ex}} \mathbf{h}  \right] \nonumber \\
	&=& \prod_{\mathclap{i=1, i\neq p}}^N   \left( 1- \mathrm{E}_{{\mathbf{h}}} \left[ \exp \left( { -\frac{ \|\tilde{\mathbf{h}}\|_2^2(1-\mu^{*})^2}{2 {\sigma_{{v}}}^2}} \right)  \mathrel{\stretchto{\mid}{4ex}} \mathbf{h}  \right]-  \mathrm{E}_{\mathbf{h}} \left[\exp \left( -\frac{\|\tilde{\mathbf{h}}\|_2^2}{\sigma_{{v}}^2} \right) \mathrel{\stretchto{\mid}{4ex}} \mathbf{h}  \right] \right).
	\end{eqnarray}
	Since $\|\tilde{\mathbf{h}}\|_2^2$ follows Chi-squared distribution with $2m$ DoF (see Lemma 2), we have
	\begin{eqnarray}
	\label{eq:a12}
	\mathrm{E}_{\tilde{\mathbf{h}}} \left[ 
	\exp{\left(-\frac{\|\tilde{\mathbf{h}}\|_2^2 }{ {\sigma_{{v}}}^2} \right)} \mathrel{\stretchto{\mid}{4ex}} \mathbf{h}  \right] 
	&=& \int_{0}^{\infty} 
	\exp{\left(-\frac{x }{  {\sigma_{{v}}}^2} \right)}  {\frac  {x^{{m-1}}\exp \left( -x\right)} {(m-1)!}} dx, \nonumber \\
	&=& \frac{1}{\left(\frac{1}{\sigma_v^2}+1\right)^m},
	\end{eqnarray}
	where the equality follows from $\int_{0}^{\infty} x^n \exp{(-ax)} dx= \frac{n!}{a^{n+1}}$ for $n=0, 1, 2, ..., a>0$.
	
	In a similar way, we have 
	\begin{eqnarray}
	\label{eq:a1311}
	\mathrm{E}_{\mathbf{h}} \left[ 
	\exp{\left(-\frac{\|\tilde{\mathbf{h}}\|_2^2(1-\mu^{*})^2}{ 2{\sigma_{{v}}}^2} \right)} \mathrel{\stretchto{\mid}{4ex}} \mathbf{h}  \right]  &=& \left(1+\frac{(1-\mu^{*})^2}{\sigma_v^2}\right)^{-m}.
	\end{eqnarray}
	
	Finally, by plugging (\ref{eq:a12}) and (\ref{eq:a1311}) into (\ref{eq:a111}), we obtain the lower bound of $\mathrm{P}({\mathcal{S}}^1)$ as 
	\begin{eqnarray}
	\label{eq:a15}
	\mathrm{P}({\mathcal{S}}^1) &=&{\mathrm{E}_{\mathbf{h}} \left[~~ \prod_{\mathclap{i=1, i\neq p}}^N \mathrm{P}\left(  \left| \Re\langle \frac{\bm{\phi}_p}{\|\bm{\phi}_p\|_2}, {\mathbf{r}^0}  \rangle\right|  \geq   \left|\Re \langle \frac{\bm{\phi}_i}{\|\bm{\phi}_i\|_2}, {\mathbf{r}^0} \rangle\right| \right)   \mathrel{\stretchto{\mid}{3ex}} \mathbf{h}  \right]}   \nonumber \\ &=& \prod_{\mathclap{i=1, i\neq p}}^N  ~\left( 1-\left(1+\frac{(1-\mu^{*})^2}{\sigma_v^2}\right)^{-m} -\left(1+\frac{1}{\sigma_v^2}\right)^{-m} \right) \nonumber \\
	&\geq& \left( 1-\left(1+\frac{(1-\mu^{*})^2}{\sigma_v^2}\right)^{-m} -\left(1+\frac{1}{\sigma_v^2}\right)^{-m} \right)^{N-1}_{\textstyle ~~~~.} 
	\end{eqnarray}
\end{IEEEproof}

We now move to the success probability for the second iteration when the first iteration is successful.
\begin{lemma}
	The probability that the support element is chosen at the second iteration under the condition that the first iteration is successful satisfies
	\begin{eqnarray}
	\mathrm{P}\left( \mathcal{S}^2 |\mathcal{S}^1\right) &\geq& \left( 1-\left(1+\frac{(1-\mu^{*})^2}{\sigma_v^2}\right)^{-m} -\left(1+\frac{1}{\sigma_v^2}\right)^{-m} \right)^{N-2}_{\textstyle ~~~~.}
	\end{eqnarray}
\end{lemma}
\begin{IEEEproof}
	When the first iteration is successful, the residual $\mathbf{r}^1$ can be expressed as   
	\begin{eqnarray}
	\label{eq:a3}
	{\mathbf{r}}^{1}&=&{\mathbf{r}^0}-\mathbf{\Phi}_{\Omega^{1}_\mathbf{s}}\hat{\mathbf{s}}^{1} \nonumber \\
	&\overset{(a)}{=}& {\mathbf{r}}^{0}-\bm{\phi}_{p}{{s}}_{p} \nonumber \\
	&=& \bm{\phi}_{q}{{s}}_{q} + \tilde{\mathbf{v}},
	\end{eqnarray}
	where $(a)$ is because the transmit symbols are known in advance ($\hat{\mathbf{s}}^1=s_p$).
	After taking similar steps to Lemma 3, one can show that $\mathrm{P}\left( \mathcal{S}^2 |\mathcal{S}^1\right)$ satisfies (we skip the detailed steps for brevity)
	%
	\begin{eqnarray}
	\mathrm{P}\left( \mathcal{S}^2 |\mathcal{S}^1\right) &=&\mathrm{P}\left( \left|\Im \langle  \frac{\bm{\phi}_q}{\|\bm{\phi}_q\|_2}, \mathbf{r}^1 \rangle\right| \geq \max_{i} \left|\Im \langle  \frac{\bm{\phi}_i}{\|\bm{\phi}_i\|_2}, \mathbf{r}^1  \rangle \right| \right) \nonumber \\
	&=& \prod_{\mathclap{i=1, i\neq p,q}}^{N} \mathrm{P}\left( \left|\Im \langle \frac{\bm{\phi}_q}{\|\bm{\phi}_q\|_2}, \mathbf{r}^1 \rangle\right| \geq  \left|\Im \langle \frac{\bm{\phi}_i}{\|\bm{\phi}_i\|_2}, \mathbf{r}^1 \rangle\right|\right) \label{eq:a4} \nonumber \\
	&\geq& \left( 1-\left(1+\frac{(1-\mu^{*})^2}{\sigma_v^2}\right)^{-m} -\left(1+\frac{1}{\sigma_v^2}\right)^{-m} \right)^{N-2}_{\textstyle ~~~~.}  \label{eq:a41}
	\end{eqnarray}
\end{IEEEproof}	

It is worth mentioning that the lower bounds of $\mathrm{P}\left( \mathcal{S}^1\right)$ and  $\mathrm{P}\left( \mathcal{S}^2 |\mathcal{S}^1\right)$ have the same form except for the exponent. We are now ready to prove the main theorem.

\begin{IEEEproof}[Proof of Theorem 1]
By combining Lemma 3 and 4, we can obtain the lower bound of the success probability $P_{succ}$ as 
\begin{eqnarray}
\label{eq:a5}
{P_{succ}}  &=& \mathrm{P}\left( \mathcal{S}^2 |\mathcal{S}^1\right) \mathrm{P}\left( \mathcal{S}^1 \right) \nonumber \\
&=& \mathrm{P}\left( \left|\Im \langle \frac{\bm{\phi}_q}{\|\bm{\phi}_q\|_2}, \mathbf{r}^1 \rangle\right| \geq \max_{i} \left|\Im \langle \frac{\bm{\phi}_i}{\|\bm{\phi}_i\|_2}, \mathbf{r}^1 \rangle \right| \right)  \mathrm{P}\left( \left|\Re \langle \frac{ \bm{\phi}_p}{\| \bm{\phi}_p\|_2}, {\mathbf{r}^0} \rangle\right| \geq \max_{i } \left| \Re \langle \frac{\bm{\phi}_i}{\|\bm{\phi}_i\|_2} ,  {\mathbf{r}^0} \rangle \right| \right) \nonumber \\ 
&=& \prod_{\mathclap{i=1, i\neq p,q}}^{N} \mathrm{P}\left( \left|\Im \langle \frac{\bm{\phi}_q}{\|\bm{\phi}_q\|_2}, \mathbf{r}^1 \rangle\right| \geq  \left|\Im \langle \frac{\bm{\phi}_i}{\|\bm{\phi}_i\|_2}, \mathbf{r}^1 \rangle\right|\right) \prod_{\mathclap{i=1, i\neq p}}^N \mathrm{P}\left( \left|\Re \langle  \frac{\bm{\phi}_p}{\|\bm{\phi}_p\|_2}, {\mathbf{r}^0}\rangle\right| \geq  \left|\Re \langle \frac{\bm{\phi}_i}{\|\bm{\phi}_i\|_2}, {\mathbf{r}^0}\rangle\right|\right)  \nonumber \\
&\geq& \left( 1-\left(1+\frac{(1-\mu^{*})^2}{\sigma_v^2}\right)^{-m} -\left(1+\frac{1}{\sigma_v^2}\right)^{-m} \right)^{(N-2)+(N-1)} \nonumber \\
&\geq& \left( 1-\left(1+\frac{(1-\mu^{*})^2}{\sigma_v^2}\right)^{-m} -\left(1+\frac{1}{\sigma_v^2}\right)^{-m} \right)^{2N}_{\textstyle ~~,}
\end{eqnarray}

which completes the proof.
\end{IEEEproof}

\addtocounter{theorem}{-3}

 Finally, we present the decoding success probability bound for general sparsity $K$.
\begin{theorem}
	The probability that the SVC-encoded packet can be successfully decoded for a given $K$ satisfies  
	\begin{eqnarray}
	P_{succ} &\gtrsim& \left( 1-\left(1+\frac{(1-\mu^{*})^2}{\sigma_v^2}\right)^{-m}  \right)^{KN}_{\textstyle ~~~.}
	\end{eqnarray}
\end{theorem}
\begin{IEEEproof}
	The success probability $P_{succ}$ is expressed as 
	\begin{eqnarray}
		{P_{succ}}  &=&\mathrm{P}\left( \mathcal{S}^1, \mathcal{S}^2, \cdots, \mathcal{S}^K  \right) \nonumber \\ &=& \mathrm{P}\left( \mathcal{S}^K |\mathcal{S}^{K-1},\cdots,\mathcal{S}^{1}\right) \cdots \mathrm{P}\left( \mathcal{S}^2 |\mathcal{S}^1\right) \mathrm{P}\left( \mathcal{S}^1 \right) \nonumber \\
		&\geq& \left( 1-\left(1+\frac{(1-\mu^{*})^2}{\sigma_v^2}\right)^{-m} -\left(1+ \frac{1}{\sigma_v^2}\right)^{-m} \right)^{(N-K)+\cdots+(N-2)+(N-1)} \nonumber \\
			&\geq& \left( 1-\left(1+\frac{(1-\mu^{*})^2}{\sigma_v^2}\right)^{-m} -\left(1+ \frac{1}{\sigma_v^2}\right)^{-m} \right)^{KN}_{\textstyle ~~~.}
	\end{eqnarray}
	Since the proof is similar to the proof of Theorem 1, we skip the detailed steps. 
\end{IEEEproof}
If $m\gg 1$, we approximately have
\begin{eqnarray}
\label{eq:4444}
{P_{succ}}  &\gtrsim& \left( 1-\left(1+\frac{(1-\mu^{*})^2}{\sigma_v^2}\right)^{-m}  \right)^{KN}_{\textstyle ~~~.}
\end{eqnarray}
It is clear from (\ref{eq:4444}) that  the decoding success probability decreases when the information vector is less sparse (i.e., $K$ is large), which matches with our expectation.

\section{Implementation Issues}
In this section, we discuss the implementation issues including codebook design, high-order modulation, diversity transmission, pilot-less transmission, and threshold selection to prevent the false alarm event.
\subsection{Codebook Design}
From our analysis in the previous section, we clearly see that a codebook with small correlation is important to improve the decoding success probability. As mentioned, as $m$ increases, the correlation between two randomly generated codewords decreases, and thus we can basically use any kind of randomly generated sequence. For example, if we use the Bernoulli random matrix, then the maximum correlation satisfies $\mu^{*} \leq \sqrt{4m^{-1}\ln\frac{N}{\delta}}$ with probability exceeding $1-\delta^2$ \cite{Joel}. 

Instead of relying on the random sequence, we can alternatively use the deterministic sequences. Well-known deterministic sequences include chirps, BCH, DFT, and second-order Reed-Muller (SORM) sequences \cite{sorm}. For example, SORM is a sequence designed to generate low correlation sequences. SORM of length $2^m$ is defined as  
\begin{equation}
\phi_{\mathbf{P},\mathbf{b}(\mathbf{a})} = \frac{(-1)^{w(\mathbf{b})}}{\sqrt{2^p}} i^{(2\mathbf{b}+\mathbf{P}\mathbf{a})^T\mathbf{a}},
\end{equation}
where $\mathbf{P}$ is a $d \times d$ binary symmetric matrix, $\mathbf{a}=[a_0 ~a_1 ~\cdots~a_{d-1}]^T$ and $\mathbf{b}=[b_0~b_1~\cdots~b_{d-1}]^T$ are binary vectors in $\mathbb{Z}_2^d$, and $w(\mathbf{b})$ is the weight (number of ones) of $\mathbf{b}$. The corresponding SORM matrix can be expressed as 
\begin{equation}
\label{eq:46}
\Phi_{rm} = \begin{bmatrix}
\mathbf{U}_{\mathbf{P}_1} & \mathbf{U}_{\mathbf{P}_2} & \cdots & \mathbf{U}_{\mathbf{P}_{2^{d(d-1)/2}}}
\end{bmatrix},
\end{equation}
where $\mathbf{U}_{\mathbf{P}_j}$ is the $2^d \times 2^d$ orthogonal matrix whose columns are the SORM sequences. The maximum correlation $\nu^{*}$ of the SORM sequence is 
\begin{equation}
\nu^{*} = \begin{cases}\frac{1}{\sqrt{2^l}}, & l=\text{rank}(\mathbf{P}_i-\mathbf{P}_j) \\ \frac{1}{\sqrt{m}}, &  l=d \end{cases}_{\textstyle .}
\end{equation}
For example, if $m=64$ and $l=d$, then $\nu^{*}=0.125$. The benefit of using SORM sequence is that the correlation between any two codewords is a constant and thus the performance variation can be minimized. 
\subsection{High-order Modulation}
Since the ensuring reliability is the top priority in URLLC, QPSK modulation would be the popular option in practice. In order to use the QPSK modulation in SVC, we set one of the nonzero entries in  $\mathbf{s}$ to 1 and the other to $j$. For example, if the nonzero positions are 5 and 7, then we set $\mathbf{s}= [0~0~0~0~1~0~j~0~0~0]^T$ and thus the transmit vector $\mathbf{x}$ can be expressed as
\begin{equation}
\label{eq:32}
\mathbf{x}=1\mathbf{c}_5+ j\mathbf{c}_7.
\end{equation}
From (\ref{eq:32}), we can easily see that elements of the transmit vector $\mathbf{x}$ are mapped to the QPSK symbol (i.e., ${x}_i \in \left\{ 1\text{+}j, 1\text{-}j, \text{-}1\text{+}j, \text{-}1\text{-}j \right\}$). {It is worth mentioning that one additional bit can be encoded by differentiating two possible choices (i.e., $[1, j]$ and $[j, 1]$).} However, this choice will increase the computational overhead of the decoding algorithm and also degrade the performance little bit. When the higher sparsity is used, this mapping can be readily extended to the high order modulation (e.g., $K=4$ for 16-QAM and $K=6$ for 64-QAM). Specifically, if $K=4$, we map the element in $\mathbf{x}$ to the 16-QAM symbol by setting two of the nonzero entries to $1$, $2$ and the remaining nonzero entries to $j$, $2j$. In a similar way, if $K=6$, then we can transmit 64-QAM symbols by setting three of the nonzero entries to $1$, $2$, $3$ and the remaining ones to $j$, $2j$, and $3j$. The normalization factor ($\alpha$ in (\ref{eq:444})) corresponding $M$-QAM is $\alpha = \sqrt{\frac{2(M-1)}{3}}$. 

\subsection{Diversity Transmission}
One can easily integrate the diversity scheme to SVC to further improve the reliability. The first option is to use the frequency diversity in which the SVC-encoded packet is repeated $L$ times in $L$ distinct frequency bands. The benefit of the frequency diversity is that the diversity gain can be achieved without increasing the transmission latency. Specifically, by applying the maximal-ratio combining at the receiver for the same symbol of the repeated packets, effective SNR can be increased and thus the BLER performance can be improved \cite{wbook}. For example, when the SVC-encoded packet is repeated for $L=8$ times, due to the power gain of the combined symbol, the required SNR to achieve the desired URLLC performance (e.g., $10^{\text{-}5}$ BLER) can be reduced from $3$ dB to $3-10\log_{10}(L)=\text{-}6$dB in AWGN environments. On top of the frequency diversity, other diversity schemes such as time, antenna, and space diversity can also be easily incorporated.
\subsection{SVC without Pilot}
When the channel is a constant or channel variation is very small (i.e., $h \approx $ const.), which is true for mobile devices under static or slowly moving environments, decoding of the SVC packet can be performed without pilot transmission, resulting in a substantial reduction of the resources, transmission power, receiver processing time, and also implementation cost. In fact, since the packet length is smaller than the channel coherence time, this assumption holds true in many realistic scenarios. Pilot-less transmission is done by slightly modifying the system model such that the system matrix equals the codebook $\mathbf{C}$ and the sparse vector contains the channel component ($\mathbf{s}^{'}=h\mathbf{s}$). That is,  
\begin{eqnarray}
\mathbf{y} &=& \mathbf{HCs+v} \nonumber \\ 
&=& \mathbf{C}\mathbf{s}^{'} +\mathbf{v}\nonumber \\
&=& 
\begin{bmatrix}
\vertbar  &  & \vertbar\\
\mathbf{c}_{1} &  \dots & \mathbf{c}_{N}\\
\vertbar &   &\vertbar\\
\end{bmatrix}
\begin{bmatrix}
h s_1\\
\vdots\\
h s_N
\end{bmatrix}
+	\begin{bmatrix}
v_1\\
\vdots\\
v_m
\end{bmatrix}_{\textstyle .}
\end{eqnarray}
Recalling that the goal of the SVC decoding is to find out the nonzero positions of $\mathbf{s}^{'}$ vector, we can perform the decoding without the channel knowledge. When the channel variation is flat in the frequency axis, tall packet structure (stretched in frequency axis) is preferred. Whereas, if the channel variation is very small in time-domain, horizontal packet (stretched in time axis) would be a good choice.
\subsection{Threshold to Prevent False Alarm Event}
\begin{figure}[t!]
	\centering
	\begin{tikzpicture}
	\begin{semilogyaxis}[
	xlabel=\footnotesize{SNR (dB)},
	ylabel=\footnotesize{Prob. of false alarm},
	xmin=-2, xmax=12,
	ymin=1e-5, ymax=1,
	grid={major},
	xtick={-2, 0, 2,  4,  6, 8, 10, 12},
	xticklabels={\footnotesize{-2},  \footnotesize{0},  \footnotesize{2}, \footnotesize{4}, \footnotesize{6},  \footnotesize{8},  \footnotesize{10},  \footnotesize{12}},
	ytick={1e-5, 1e-4 ,1e-3, 1e-2, 1e-1, 1},
	yticklabels={\footnotesize{$10^{-5}$}, \footnotesize{$10^{-4}$}, \footnotesize{$10^{-3}$}, \footnotesize{$10^{-2}$}, \footnotesize{$10^{-1}$}, \footnotesize{$10^{0}$}},
	legend cell align=left,
	legend style={legend pos=south west},
	scale=1.0
	]
	\addplot[color=red,mark=square, mark options={solid}] coordinates {
		(-2,1 ) (-1.5, 0.961) (-1.1, 0.856) (0.6, 0.7270) (1.2, 0.621) (3, 0.2184)(5, 0.052) (7, 0.011) (9,  1.4000e-03) (12, 2.0000e-05) (14, 3e-6)
	};
		\addplot[dashed, color=black,mark=o] coordinates {
		(-2, 0.8) (0, 0.8) (2, 0.6) (3, 0.4) (4.6, 1.2e-1)  (7.2, 1e-2) (9.2, 1e-3) (10.6, 1e-4) (11.4, 1e-5)
	};
	\legend{ \scriptsize{With residual threshold}, \scriptsize{With 16-bit CRC}}
	\end{semilogyaxis}
	\end{tikzpicture}
	\caption{Decoding failure as a function of SNR ($P_{th}=10^{-5}$).}
	\label{fig:result9}
\end{figure}
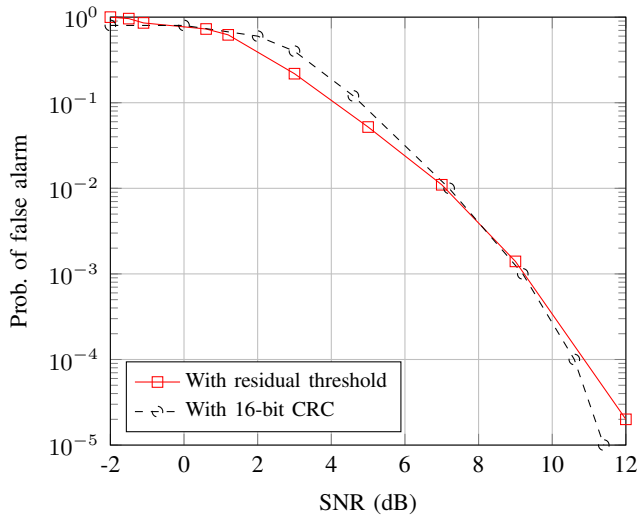

To distinguish the false alarm event from the normal decoding process, we examine the probability that the residual after the sparse recovery algorithm is not pure noise. In fact, if the SVC decoding is finished successfully, the residual contains the noise contribution only ($\mathbf{r}^K=\mathbf{v}$) so that the residual power $\|\mathbf{r}^K\|^2_2$ can be readily modeled as a Chi-squared random variable with $2m$ degree of freedom. Naturally, one can reject this hypothesis if the residual power is too large and lies outside of the pre-defined confidence interval. In other words, if $\|\mathbf{r}^K\|^2_2 > F^{-1}_{\|\mathbf{v}\|^2_2}(1-P_{th})$ where $P_{th}$ is the pre-defined probability threshold (e.g., $P_{th}= 0.01$) and $F^{-1}_{\|\mathbf{v}\|^2_2}$ is the inverse cumulative distribution function of Chi-squared random variable, then we declare the hypothesis is not true (i.e., decoding is not successful) and discard the decoded packet. To evaluate the effectiveness of this thresholding approach, we simulate the probability of false alarm as a function of SNR for the conventional 16-bit CRC and the proposed residual-based thresholding. As is clear from Fig. \ref{fig:result9}, the residual-based thresholding performs similarly to the CRC-based error checking. 


\section{Simulations and Discussions}
\subsection{Simulation Setup}
In this section, we examine the performance of the proposed SVC technique. Our simulation setup is based on the downlink OFDM system in the 3GPP LTE-Advanced Rel.13 \cite{lte}. As a channel model, we use AWGN and realistic ITU channel models including extended typical urban (ETU) and extended pedestrian-A (EPA) channel model \cite{lte}. For comparison, we also investigate the performance of the conventional PDCCH of LTE-Advanced system, polar code-based PDCCH of 5G systems \cite{polar}, and AWGN lower bound. We test the transmission of $b$ bit information which consists of information bit $b_i$ and CRC bit $b_c$. In the conventional PDCCH method, the convolution code with rate $\frac{1}{3}$ with the $16$-bit CRC is employed. Since the block size of the polar code is not flexible, we set the rate $\frac{1}{4}$ to test similar conditions ($b=2^4$ and $m=32$). In the proposed SVC algorithm, we set the random binary spreading codebook with $N=96$ and $K=2$. To ensure the fair comparison, we use the same number of resources ($m=42$ with $L=8$ repetitions) in the control packet transmission. As a performance measure, we use BLER of the code blocks.
\begin{figure}[t!]
	\centering
			\begin{tikzpicture}
		\begin{semilogyaxis}[
		xlabel=\footnotesize{SNR (dB)},
		ylabel=\footnotesize{BLER},
		xmin=-12, xmax=0,
		ymin=10e-6, ymax=1,
		grid={both}, yminorticks=true, minor y tick num=9, 
		xtick={-12, -10,   -8  , -6, -4, -2, 0},
		xticklabels={\footnotesize{-12}, \footnotesize{-10}, \footnotesize{-8},  \footnotesize{-6} , \footnotesize{-4}, \footnotesize{-2} ,\footnotesize{0}},
		ytick={10e-6, 10e-5, 10e-4, 10e-3, 10e-2,10e-1, 1},
		yticklabels={\footnotesize{$10^{-5}$}, \footnotesize{$10^{-4}$}, \footnotesize{$10^{-3}$}, \footnotesize{$10^{-2}$}, \footnotesize{$10^{-1}$}, \footnotesize{$1$}},
		legend cell align=left,
		legend style={legend pos=north east},
		scale=0.94
		]
		\addplot[color=blue] coordinates {
			(-1.46580802041939-9, 1.00000000000000e-06)
			(-1.46586028028238-9, 1.59752725346392e-06)
			(-1.46594088443954-9, 2.55209332555996e-06)
			(-1.46606504790020-9,	4.07703864096540e-06)
			(-1.46625604954563-9, 6.51318034236771e-06)
			(-1.46654943803498-9, 1.04049831036579e-05)
			(-1.46699938683487-9, 1.66222440799250e-05)
			(-1.46768826753951-9,	2.65544879314094e-05)
			(-1.46874102422759-9, 4.24215181722052e-05)
			(-1.47034668114137-9, 6.77695314134125e-05)
			(-1.47279040481839-9, 0.000108263673387405)
			(-1.47650111802987-9, 0.000172954168796496)
			(-1.48212194192189-9, 0.000276298998252602)
			(-1.49061404752073-9, 0.000441395179813309)
			(-1.50340933552289-9, 0.000705140829299368)
			(-1.52263458199364-9, 0.00112648169233589)
			(-1.55144080631253-9, 0.00179958520403473)
			(-1.59448954424627-9, 0.00287488640837591)
			(-1.65867834006305-9,	0.00459270938799350)
			(	-1.75424368874085-9,	0.00733697841455921)
			(-1.89648912049432-9,	0.0117210229753348)
			(-2.10861656784624-9,	0.0187246536415741)
			(-2.42666473293131-9, 0.0299131445040870)
			(-2.90887785174526-9, 	0.0477870635820833)
			(-3.65558399853535-9, 0.0763411364353911)
			(-4.85838491308664-9, 0.121957046015944)
			(-6.95353754966826-9, 0.194829704762424)
			(-11.3503705783626-9, 0.311245763142300)
			(-48.1200002613451-9,	0.497223589145000) 
		};
		\addplot[color=red,mark=triangle, mark options={solid}] coordinates {
			(-13, 0.25432)
			(-12, 0.18432)
			(-11, 0.09432)
			(-10, 0.027173)
			(-9, 0.006239)
			(-8, 0.000889)
			(-7, 0.000104)
			(-6, 1e-05)
			(-5, 1e-06)
			(2, 2.12e-07)
		};
		\addplot[color=black,mark=x, mark options={solid}] coordinates {
			(-12, 0.5555)
			(-11, 0.46655)
			(-10, 0.34655)
			(-9, 0.204)
			(-8, 0.096503)
			(-7, 0.035716)
			(-6, 0.009792)
			(-5, 0.001926)
			(-4, 0.000266)
			(-3, 2.3e-05)
			(-2, 1.3e-06)
		};
		\addplot[color=blue,mark=o, mark options={solid}] coordinates {
			(-11.45-2, 0.684) (-11.45-1.5, 0.597)	(-11.45-1, 0.523) 	(-11.45-0.5, 0.429)
			(-11.45, 0.355) (-11.45+0.4, 0.284) (-11.45+0.8, 0.22) (-11.45+1.2, 0.1569)
			(-9.45, 0.0734) (-9.45+1, 0.021)  (-9.45+1.5, 0.0082) (-9.45+2.5, 0.0012) (-9.45+3.5, 0.00014)
			(-9.45+4.5, 0.000012)
		};
		\legend{\scriptsize {Lower bound},  \scriptsize{SVC}, \scriptsize{PDCCH},  \scriptsize{Polar code}}
		\end{semilogyaxis}
		\end{tikzpicture}
%
%
				\begin{tikzpicture}
		\begin{semilogyaxis}[
		xlabel=\footnotesize{SNR (dB)},
		ylabel=\footnotesize{BLER},
		xmin=-5, xmax=35,
		ymin=10e-6, ymax=1,
		grid={major},
		xtick={-15, -10,   -5  , 0, 5, 10, 15, 20, 25, 30, 35},
		xticklabels={\footnotesize{-15}, \footnotesize{-10}, \footnotesize{-5},  \footnotesize{0} , \footnotesize{5}, \footnotesize{10} ,\footnotesize{15}, \footnotesize{20}, \footnotesize{25}, \footnotesize{30}, \footnotesize{35}},
		ytick={10e-6, 10e-5, 10e-4, 10e-3, 10e-2,10e-1, 1},
		yticklabels={\footnotesize{$10^{-5}$}, \footnotesize{$10^{-4}$}, \footnotesize{$10^{-3}$}, \footnotesize{$10^{-2}$}, \footnotesize{$10^{-1}$}, \footnotesize{$1$}},
		legend cell align=left,
		legend style={legend pos=north east},
		scale=0.94
		]
			\addplot[dashed, color=black,mark=x, mark options={solid}] coordinates {
			(-10, 0.8746)
			(-8,  0.76425)
			(-6, 0.60745)
			(-4, 0.44285)
			(-2, 0.2863)
			(0, 0.1638)
			(2, 0.08305)
			(4, 0.03835)
			(6, 0.01845)
			(8, 0.00795)
			(10, 0.0037)
			(12, 0.0017)
			(14, 0.00057)
			(16, 0.00019)
			(18, 0.00006)
			(20, 0.00002)
			(22, 0.000005)
		};
		\addplot[dashed, color=red,mark=triangle, mark options={solid}] coordinates {
			(-10, 0.7051)
			(-8,  0.5286)
			(-6, 0.3509)
			(-4, 0.2073)
			(-2, 0.10185)
			(0, 0.0477)
			(2, 0.02195)
			(4, 0.01015)
			(6, 0.0041)
			(8, 0.0017)
			(10, 0.00065)
			(12, 0.0002)
			(14, 0.000065)
			(16, 0.000018)
			(18, 0.000005)
		};
		\addplot[ color=black,mark=x, mark options={solid}] coordinates {
			(-10, 0.5446)
			(-8,  0.46425)
			(-6, 0.28745)
			(-4, 0.2285)
			(-2, 0.1563)
			(0, 0.10842)
			(2, 0.069505)
			(4, 0.043846)
			(6, 0.027982)
			(8, 0.017822)
			(10, 0.010622)
			(12, 0.0065468)
			(14, 0.0033348)
			(16, 0.0019432)
			(18, 0.0012401)
			(20, 0.00065283)
			(22, 0.00037785)
			(24, 0.00017855)
			(26, 0.00009123)
			(28, 0.00005049)
			(30, 0.0000223)
			(32, 0.00001)
		};
		\addplot[ color=red,mark=triangle, mark options={solid}] coordinates {
			(-10, 0.3946)
			(-8,  0.28425)
			(-6, 0.1945)
			(-4, 0.1185)
			(-2, 0.07563)
			(0, 0.055584)
			(2, 0.035128)
			(4, 0.022586)
			(6, 0.013951)
			(8, 0.0085948)
			(10, 0.0046199)
			(12, 0.0026702)
			(14, 0.0012826)
			(16, 0.0007733)
			(18, 0.0003551)
			(20, 0.00019783)
			(22, 8.923e-05)
			(24, 3.968e-05)
			(26, 1.668e-05)
			(28, 0.668e-05)
		};
		\legend{\scriptsize{PDCCH (ETU channel)},  \scriptsize{SVC (ETU channel)}, \scriptsize{PDCCH (EPA channel)},  \scriptsize{SVC  (EPA channel)}}
		\end{semilogyaxis}
		\end{tikzpicture}
		
		\footnotesize{~~~~~~~~(a)~~~~~~~~~~~~~~~~~~~~~~~~~~~~~~~~~~~~~~~~~~~~~~~~~~~~~~~~~~~~~~~~~~~~~~~(b)}
		\caption{BLER performance as a function of SNR ($b_i=12$, $b_c=16$, $m=42$, $L=8$, and $N=96$) for (a) AWGN channel and (b) ETU and EPA channel.}
		\label{fig:result2b}
\end{figure}
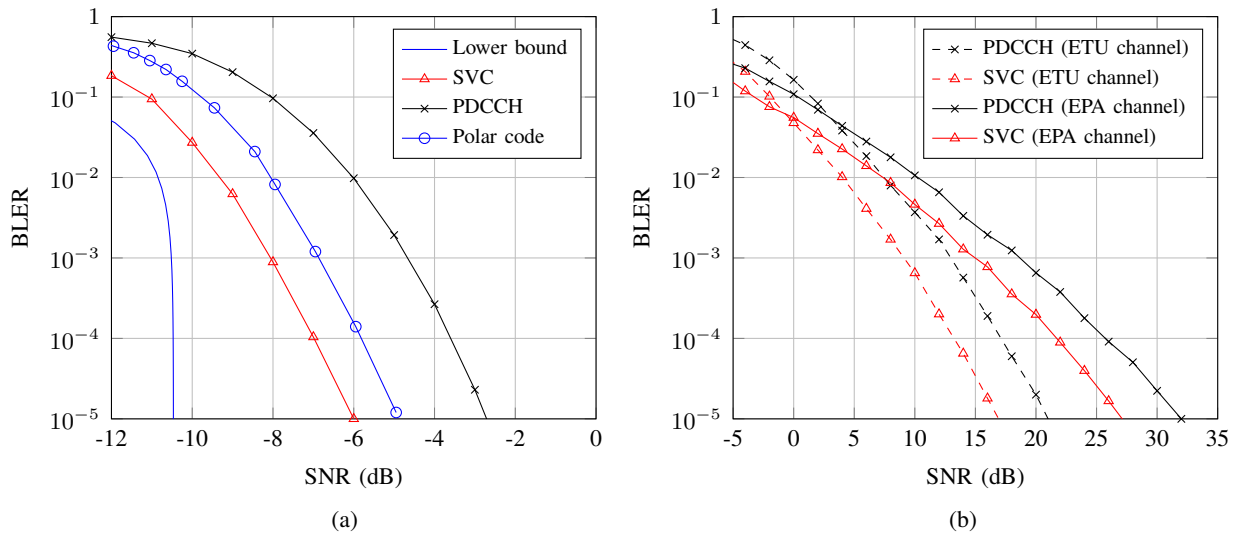
\subsection{Simulation Results}
In Fig. \ref{fig:result2b}(a), we investigate the BLER performance of the proposed SVC method and competing schemes under AWGN channel condition. We observe that the proposed SVC technique outperforms the conventional PDCCH and polar code-based scheme, achieving more than 4 dB gain over the conventional PDCCH and about 1.1 dB gain over the polar code-based scheme at $10^{\text{-}5}$ BLER point. Even in realistic scenarios such as EPA and EVA channels in LTE-Advanced, we observe that the performance gain of the proposed SVC scheme over competing schemes is maintained  (see Fig. \ref{fig:result2b}(b)). 

\begin{figure}[t!]
	\centering
	\begin{tikzpicture}
	\begin{semilogyaxis}[
	xlabel=\footnotesize{SNR (dB)},
	ylabel=\footnotesize{BLER},
	xmin=-12, xmax=8,
	ymin=10e-6, ymax=1,
	grid={major},
	xtick={-12,  -8,    -4,  0, 4, 8},
	xticklabels={\footnotesize{-12},  \footnotesize{-8},    \footnotesize{-4},  \footnotesize{0}, \footnotesize{4}, \footnotesize{8}},
	ytick={10e-6, 10e-5, 10e-4, 10e-3, 10e-2,10e-1, 1},
	yticklabels={\footnotesize{$10^{-5}$}, \footnotesize{$10^{-4}$}, \footnotesize{$10^{-3}$}, \footnotesize{$10^{-2}$}, \footnotesize{$10^{-1}$}, \footnotesize{$1$}},
	legend cell align=left,
	legend style={legend pos=north east,legend columns=2},
	scale=1
	]
	\addplot[dashed, color=black,mark=o, mark options={solid}] coordinates {
		(-12, 0.5555)
		(-11, 0.46655)
		(-10, 0.34655)
		(-9, 0.204)
		(-8, 0.096503)
		(-7, 0.035716)
		(-6, 0.009792)
		(-5, 0.001926)
		(-4, 0.000266)
		(-3, 2.3e-05)
		(-2, 1.3e-06)
	};
	\addplot[color=red,mark=o, mark options={solid}] coordinates {
		(-13, 0.25432)
		(-12, 0.18432)
		(-11, 0.09432)
		(-10, 0.027173)
		(-9, 0.006239)
		(-8, 0.000889)
		(-7, 0.000104)
		(-6, 1e-05)
		(-5, 1e-06)
		(2, 2.12e-07)
	};
	\addplot[dashed, color=black,mark=square, mark options={solid}] coordinates {
		(-10.5, 0.97595)
		(-8.5, 0.75908)
		(-6.5, 0.25846)
		(-4.5, 0.025749)
		(-3.5, 0.002832)
		(-2.5, 0.00025409)
		(-0.5, 0.0000022)
	};
	\addplot[color=red,mark=square, mark options={solid}] coordinates {
		(-12, 0.9091)
		(-10, 0.5970)
		(-8, 0.1686)
		(-6, 0.0119)
		(-4, 1.8459e-04)
		(-2, 8.234e-07)
	};
	\addplot[dashed, color=black,mark=triangle, mark options={solid}] coordinates {
		(-12, 0.99992)
		(-10, 0.99892)
		(-8, 0.97595)
		(-6, 0.75908)
		(-4, 0.25846)
		(-2, 0.025749)
		(-1, 0.002832)
		(0, 0.00025409)
		(2, 0.0000022)
	};
	\addplot[color=red,mark=triangle, mark options={solid}] coordinates {
		(-12, 1)
		(-10, 0.9988)
		(-8, 0.9553)
		(-6, 0.5982)
		(-4, 0.1097)
		(-2, 0.0041)
		(0, 4.4279e-5)
		(1, 2.12e-06)
	};
	\addplot[dashed, color=black,mark=diamond, mark options={solid}] coordinates {
		(-10, 1.0)
		(-8, 0.99976)
		(-6, 0.9687)
		(-2, 0.56932)
		(0, 0.080938)
		(2, 0.0022035)
		(4, 0.000012035)
		(6, 0.00000062035)
	};
	\addplot[color=red,mark=diamond, mark options={solid}] coordinates {
		(-8, 1)
		(-4, 1)
		(-3, 0.8469)
		(-2, 0.5224)
		(-1, 0.2045)
		(0, 0.0480)
		(1, 0.0076)
		(2, 6.478e-04)
		(3, 4e-05)
		(4, 1.5e-06)
	};
	\legend{\scriptsize{PDCCH ($b_i=12$)}, \scriptsize{SVC ($b_i=12$)},  \scriptsize{PDCCH ($b_i=24$)}, \scriptsize{SVC ($b_i=24$)}, \scriptsize{PDCCH ($b_i=48$)},  \scriptsize{SVC ($b_i=48$)}, \scriptsize{PDCCH ($b_i=96$)},\scriptsize{SVC ($b_i=96$)}}
	\end{semilogyaxis}
	\end{tikzpicture}
	\caption{BLER performance for various size of control information ($L=8$).}
	\label{fig:result3}
\end{figure}
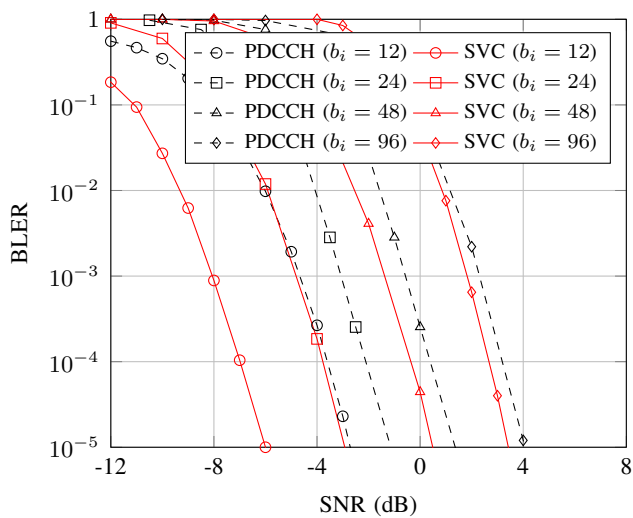
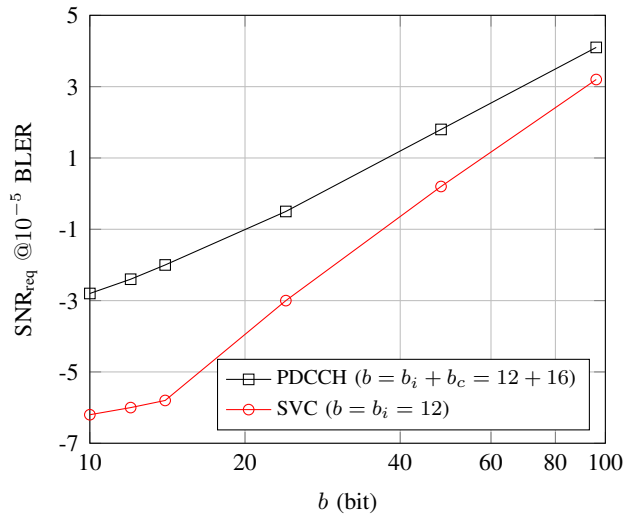
\begin{figure}[t!]
	\centering
	\begin{tikzpicture}
	\begin{semilogxaxis}[
	xlabel=\footnotesize{$b$ (bit)},
	ylabel=\footnotesize{$\text{SNR}_{\text{req}}~ @ 10^{-5}$ BLER},
	xmin=10, xmax=100,
	ymin=-7, ymax=5,
	grid={major},
	xtick={10, 20, 40, 60, 80, 100},
	xticklabels={\footnotesize{10}, \footnotesize{20}, \footnotesize{40}, \footnotesize{60}, \footnotesize{80}, \footnotesize{100}},
	ytick={-7, -5, -3, -1, 1, 3, 5},
	yticklabels={\footnotesize{-7}, \footnotesize{-5}, \footnotesize{-3}, \footnotesize{-1}, \footnotesize{1}, \footnotesize{3}, \footnotesize{5}},
	legend cell align=left,
	legend style={legend pos=south east},
	scale=1
	]
	\addplot[color=black,mark=square, mark options={solid}] coordinates {
		(10, -2.8)
		(12,-2.4)
		(14, -2)
		(24, -0.5)
		(48, 1.8)
		(96, 4.1)
	};
	\addplot[color=red,mark=o, mark options={solid}] coordinates {
		(10, -6.2)
		(12,-6)
		(14, -5.8)
		(24, -3)
		(48, 0.2)
		(96, 3.2)
	};
	\legend{\scriptsize{PDCCH ($b=b_i+b_c=12+16$)},\scriptsize{SVC  ($b=b_i=12$)}}
	\end{semilogxaxis}
	\end{tikzpicture}
	\caption{Minimum required SNR for achieving $10^{\text{-}5}$ BLER ($m=42$, $L=8$, and $N=96$)}
	\label{fig:result4}
\end{figure}
\begin{figure}
		\centering
	\begin{tikzpicture}
\begin{axis}[
scale=1,
ybar,
axis line style = { opacity = 1},
tickwidth         = 0pt,
enlarge y limits  = 0,
enlarge x limits  = 0.1,
symbolic x coords = {0.21 ms ($n=1$), 0.42 ms ($n=2$), 0.64 ms ($n=3$), 0.85 ms ($n=4$), 1.07 ms ($n=5$), 1.28 ms ($n=6$)},
ymin=0,
ymax=1,
xtick=data,
x tick label style={text width=1.1cm,align=center},
grid=major,
ylabel={Probability},
xlabel=\footnotesize{Transmission latency (ms)},
]
\addplot[pattern=north west lines, pattern color=blue] coordinates {     (0.21 ms ($n=1$), 0.8156)  (0.42 ms ($n=2$), 0.179) (0.64 ms ($n=3$), 0.005) (0.85 ms ($n=4$), 4.45e-6) (1.07 ms ($n=5$), 0.000) (1.28 ms ($n=6$), 0) };
\addplot coordinates {     (0.21 ms ($n=1$), 0.4445)  (0.42 ms ($n=2$), 0.3629) (0.64 ms ($n=3$), 0.1739) (0.85 ms ($n=4$), 0.01839) (1.07 ms ($n=5$), 0.00018) (1.28 ms ($n=6$), 1.78129e-06)};
\legend{\scriptsize{SVC}, \scriptsize{PDCCH}}
\end{axis}
	\end{tikzpicture}
	\caption{Probability of transmission latency for achieving $10^{\text{-}5}$ BLER ($b_i=12$, $m=42$, $L=8$, $N=96$, and SNR=$-12$ dB).}
	\label{fig:resulta}
\end{figure}
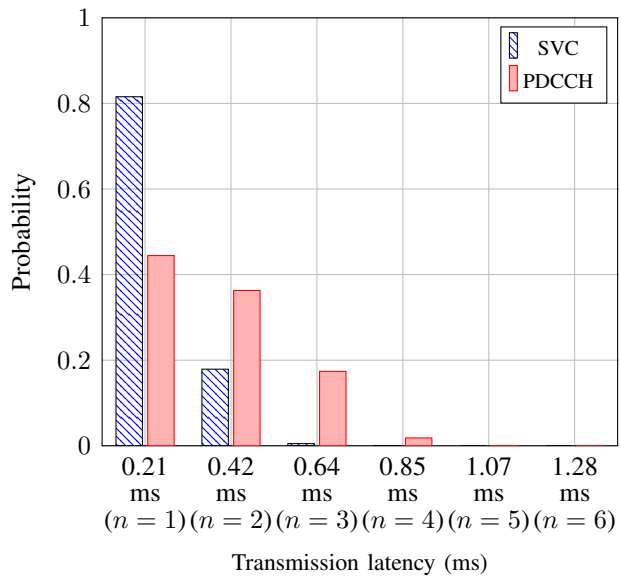
\begin{figure}[t!]
	\centering
	\begin{tikzpicture}
	\begin{semilogyaxis}[
	xlabel=\footnotesize{SINR (dB)},
	ylabel=\footnotesize{BLER},
	xmin=-12, xmax=3,
	ymin=10e-6, ymax=1,
	grid={major},
	xtick={-12, -9, -6,  -3,  0 , 3},
	xticklabels={\footnotesize{-12},  \footnotesize{-9},  \footnotesize{-6}, \footnotesize{-3},  \footnotesize{0},  \footnotesize{3} },
	ytick={10e-6, 10e-5, 10e-4, 10e-3, 10e-2,10e-1, 1},
	yticklabels={\footnotesize{$10^{-5}$}, \footnotesize{$10^{-4}$}, \footnotesize{$10^{-3}$}, \footnotesize{$10^{-2}$}, \footnotesize{$10^{-1}$}, \footnotesize{$1$}},
	legend cell align=left,
	legend style={legend pos=south west},
	scale=1
	]
	\addplot[dashed, color=black,mark=square, mark options={solid}] coordinates {
		(-12, 0.5555)
		(-11, 0.46655)
		(-10, 0.34655)
		(-9, 0.204)
		(-8, 0.096503)
		(-7, 0.035716)
		(-6, 0.009792)
		(-5, 0.001926)
		(-4, 0.000266)
		(-3, 2.3e-05)
		(-2, 1.3e-06)
	};
	\addplot[dashed, color=black,mark=o] coordinates {
		(-12.1319, 0.83286)
		(-10.2119, 0.71883)
		(-8.3312, 0.58898)
		(-6.5138, 0.47156)
		(-4.7814, 0.38312)
		(-3.19, 0.31975)
		(-1.76, 0.279)
		(-0.5354, 0.25186)
		(0.4667, 0.23567)
		(1.2445, 0.22565)
		(1.8145, 0.21859)
		(2.2185, 0.21399)
		(2.4942, 0.21199)
	};
	\addplot[color=red,mark=square, mark options={solid}] coordinates {
		(-13, 0.25432)
		(-12, 0.18432)
		(-11, 0.09432)
		(-10, 0.027173)
		(-9, 0.006239)
		(-8, 0.000889)
		(-7, 0.000104)
		(-6, 1e-05)
		(-5, 1e-06)
		(2, 2.12e-07)
	};
	\addplot[color=red,mark=o] coordinates {
		(-12.1319, 0.36301)
		(-10.2119, 0.13389)
		(-8.3312, 0.032453)
		(-6.5138, 0.0051511)
		(-4.7814, 0.00087343)
		(-3.19, 0.000111496)
		(-1.76, 1.561e-05)
		(-0.5354, 7.11e-06)
	};
	\legend{ \scriptsize{PDCCH without interf.}, \scriptsize{PDCCH with interf.},  \scriptsize{SVC without interf.},  \scriptsize{SVC with interf.}}
	\end{semilogyaxis}
	\end{tikzpicture}
	\caption{BLER performance as a function of SINR ($b_i=12$, $m=42$, $L=8$, $N=96$, and interference power is half of the signal power).}
	\label{fig:result5}
\end{figure}
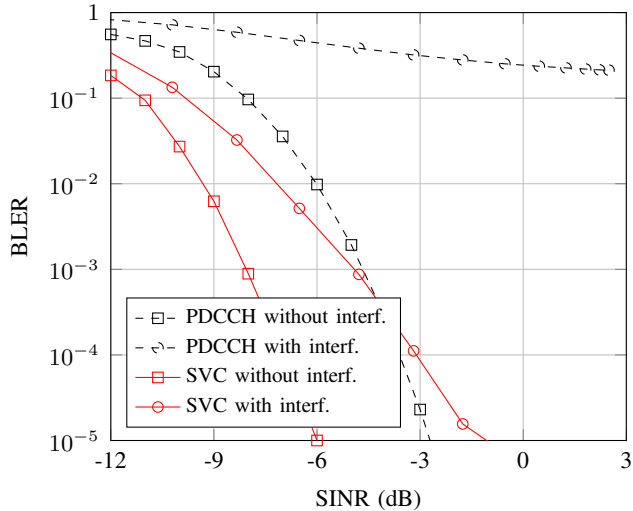

In Fig. \ref{fig:result3}, we evaluate the BLER performance of PDCCH and SVC as a function of SNR for various information bit size ($b_i=12, 24, 48,$ and $96$). These results demonstrate that the proposed SVC technique can deliver more information bits than the conventional PDCCH can support. For example, SVC can deliver twice more information than PDCCH in the low SNR region (for example, $b_i=12$ of PDCCH and $b_i=24$ of SVC in Fig. \ref{fig:result3}). To further investigate this, we plot the minimum SNR to achieve the target BLER as a function of the information bit size in Fig. \ref{fig:result4}. For example, to achieve $10^{\text{-}5}$ BLER with $b=10$, it requires $\text{-}2.9$ dB for PDCCH while $\text{-}6.2$ dB SNR for SVC, resulting in 3.3 dB gain in performance. It is worth mentioning that the coding gain of the conventional PDCCH improves with the codeblock size so that the gap between the SVC and PDCCH diminishes gradually as the number of information bits increases. 

Next, we evaluate the latency performance of the SVC and PDCCH (see Fig. \ref{fig:resulta}). In this experiments, we plot the distribution of transmission latency to achieve $10^{\text{-}5}$ BLER when $n$-repetition scheme is employed. Transmission latency is defined as the time from the initial transmission to the time that the packet is successfully decoded at the mobile terminal.\footnote{In our experiments, we ignored the decoding latency.} These results demonstrate that most of SVC packets satisfy the URLLC requirement (1 ms latency).

Finally, we investigate the performance of SVC in the small cell scenarios where the received signal contains a considerable amount of interference from adjacent basestations. Note that densely deployed small cell (pico, femto, and micro) environments will play a key role to enhance the cell throughput in 5G and how to manage the interference is the key to the success of small cell networks. In our simulations, we set the power level of interference to half of the desired cell signal. Since the SVC transmission is based on the multi-code spreading and also the effective transmit power per symbol is large (see Section III.A), SVC can effectively manage the interference. Whereas, since the conventional PDCCH has no such interference protection mechanism, error correction capability of PDCCH is degraded significantly and thus the PDCCH performs very poor as shown in Fig. \ref{fig:result5}.
\section{Conclusion}
In this paper, we have proposed the short packet transmission strategy for URLLC. The key idea behind the proposed SVC technique is to transform an information vector into the sparse vector in the transmitter and to exploit the sparse recovery algorithm in the receiver. Metaphorically, SVC can be thought as a marking dots to the empty table. As long as the number of dots is small enough and the measurements contain enough information to figure out the marked cell positions, accurate decoding of SVC packet can be guaranteed. We demonstrated from the numerical evaluations that the proposed SVC scheme is very effective in URLLC scenarios. In this paper, we restricted our attention to the URLLC scenario but we believe that there are many other applications such as mMTC that the SVC technique can be applied to. Also, there are many interesting extensions and variations worth investigating, such as the information embedding in nonzero positions, channel aware sparse vector coding, and combination of SVC and error correction codes.
\begin{appendices}
{\section{Proof of Lemma 1}
\setcounter{equation}{0}
\renewcommand{\theequation}{\thesection.\arabic{equation}}	
Let $\mathbf{u}_i = \frac{\mathbf{a}_i}{\| \mathbf{a}_i \|_2}$, then it is clear that $\mathbf{u}_i$ is a random vector with zero mean and unit variance. Also, let $X=\frac{\mathbf{a}_i^T \mathbf{a}_j }{\|\mathbf{a}_i\|_2}$, then $X=\mathbf{u}_i^T\mathbf{a}_j$. One can easily show that $X$ conditioned on any realization of $\mathbf{u}_i=u$ is a standard Gaussian. This is because $E[X|\mathbf{u}_i=u]=E[u^T\mathbf{a}_j]=u^TE[\mathbf{a}_j]=0$ and $Var(X|\mathbf{u}_i=u)=E[u^T\mathbf{a}_j\mathbf{a}_j^Tu]=u^Tu=1$. Further,
\begin{eqnarray}
f_X(x) &=&  \int_u f_{X|\mathbf{u}_i}(x|u) f_{\mathbf{u}_i}(u)du \nonumber \\ 
&=& \frac{1}{\sqrt{2\pi}} \exp \left(-\frac{x^2}{2}\right) \int_u f_{\mathbf{u}_i}(u)du \nonumber \\
&=& \frac{1}{\sqrt{2\pi}} \exp \left(-\frac{x^2}{2}\right), 
\end{eqnarray}
which is the desired result.

}
{\section{Derivation of (\ref{eq:a66})}
	\setcounter{equation}{0}
	\renewcommand{\theequation}{\thesection.\arabic{equation}}	
Noting that {$\bm{\phi}_i= [\tilde{h}_{11}{c}_{1i}~\tilde{h}_{22}{c}_{2i}~\cdots~\tilde{h}_{mm}{c}_{mi}]^T$} and $\mu_{ij} = \frac{\bm{\phi}_i^T \bm{\phi}_j}{\|\tilde{\mathbf{h}}\|_2^2 }$, we have 
	\begin{eqnarray}
	\langle \frac{\bm{\phi}_i}{\| \bm{\phi}_i \|_2}, \bm{\phi}_j \rangle &=& \frac{\bm{\phi}_i^T \bm{\phi}_j}{\|\bm{\phi}_i\|_2}. 
		\end{eqnarray}
Since $\|\bm{\phi}_i\|_2 = \sqrt{|\tilde{h}_{11} c_{1i}|^2 +  \cdots + |\tilde{h}_{mm}c_{mi}|^2 } =  \|\tilde{\mathbf{h}}\|_2 $, we have		
		\begin{eqnarray}
			\label{eq:c1}
	\langle \frac{\bm{\phi}_i}{\| \bm{\phi}_i \|_2}, \bm{\phi}_j \rangle 
	&=& \|\tilde{\mathbf{h}}\|_2 \frac{\bm{\phi}_i^T \bm{\phi}_j}{\|\tilde{\mathbf{h}}\|_2^2 } \nonumber \\	
	&=& \|\tilde{\mathbf{h}}\|_2 \mu_{ij}.
	\end{eqnarray}
	In particular, $i=j$,  $\mu_{ij}=1$ and thus 
\begin{eqnarray}
\label{eq:c3}
\langle \frac{\bm{\phi}_i}{\|\bm{\phi}_i\|_2} , \bm{\phi}_i \rangle &=& \|\tilde{\mathbf{h}}\|_2.
\end{eqnarray}
From (\ref{eq:c1}) and (\ref{eq:c3}), we have  $\langle \frac{\bm{\phi}_i}{\|\bm{\phi}_i\|_2} , \bm{\phi}_j \rangle  = \begin{cases} \|\tilde{\mathbf{h}}\|_2 & \text{for  $i = j$} \\   \|\tilde{\mathbf{h}}\|_2 \mu_{ij} & \text{for $i \neq j.$}  \end{cases}$
}

\end{appendices}




\begin{thebibliography}{1}
	\bibitem{ITA} H. Ji, S. Kim, and B. Shim, ``Sparse Vector Coding for Ultra Short Packet	Transmission,''  {\it in Proc. Inform. Theory and Appl. (ITA) Workshop}, Feb. 11-16, 2018.
	\bibitem{ITU-R} Rec. ITU-R M.2083-0, ``IMT Vision - Framework and Overall Objectives of the Future Development of IMT for 2020 and Beyond,'' Sep, 2015.
	\bibitem{MIoT} P. Schulz, M. Matth\'{e}, H. Klessig, M. Simsek, G. Fettweis, J. Ansari, S. A. Ashraf, B. Almeroth,
	J. Voigt, I. Riedel, A. Puschmann, A. Mitschele-Thiel, M. M\"{u}ller, T. Elste, and M. Windisch, ``Latency Critical IoT Applications in 5G:
	Perspective on the Design of Radio Interface and Network Architecture,'' {\it IEEE Commun. Mag.}, vol. 55, no. 2, pp. 70-78, 2017.
	\bibitem{SI} 3GPP Technical Report 38.802, ``Study on New Radio Access Technology Physical Layer Aspects (Release 14),'' v14.1.0, 2017.
	\bibitem{noma} C. Bockelmann, N. Pratas, H. Nikopour, K. Au, T. Svensson, C. Stefanovic, and A. Dekorsy, ``Massive Machine-type Communications in 5G: Physical and MAC-layer solutions,'' {\it IEEE Commun. Mag.}, vol. 54, no. 9, pp. 59-65, 2016.
	\bibitem{urllc} H. Ji, S. Park, J. Yeo, Y. Kim, J. Lee, and B. Shim, ``Ultra Reliable and Low Latency Communications in 5G: Physical Layer Aspects,'' {\it IEEE Wireless Commun.}, vol. 26, no.2, pp.100-107, June, 2018.
	\bibitem{short} B. Lee, S. Park, D. Love, H. Ji, and B. Shim, ``Packet Structure and Receiver Design for Low Latency Wireless Communications with Ultra Short Packets,'' {\it IEEE Trans. on Commun.}, vol. 66, no. 2, pp. 796-807, Feb. 2018.
	\bibitem{DelayBudget} O. Yilmaz, Y. Wang, N. A. Johansson, N. Brahmi, S. A. Ashraf, and Joachim Sachs, ``Analysis Ultra-Reliable and Low-Latency 5G Communication for a Factory Automation Use Case," {\it In Proc. IEEE Int. Conf. on Comm. (ICC) Workshop}, pp. 1190 - 1195, 2015.
	\bibitem{Div} N. A. Johansson, Y.Wang, E. Eriksson, and M. Hessler, ``Radio Access for Ultra-Reliable and Low-Latency 5G Communications," {\it In Proc. IEEE Int. Conf. on Comm. (ICC) Workshop}, pp. 1185 - 1189, 2015. 
	\bibitem{TR38913} 3GPP Technical Report 38.913, ``Study on Scenarios and Requirements for Next Generation Access Technologies (Release 14),'' v14.2.0, 2017.
	\bibitem{lte} S. Sesia, M. Baker, and I. Toufik, {\it LTE-the UMTS Long Term Evolution: from Theory to Practice,} John Wiley \& Sons., 2012.
	\bibitem{tip} J. W. Choi, B. Shim, Y. Ding, B. Rao, and D. I. Kim, ``Compressed Sensing for Wireless Communications: Useful Tips and Tricks.,'' {\it  IEEE Commun. Survey and Tutorials,}, vol. 19, pp. 1527-1550, 2017.
	\bibitem{Donoho} D. L. Donoho, ``Compressed Sensing,'' {\it IEEE Trans. on Inform. Theory}, vol.52, no. 4, pp. 1289-1306, 2006.
	\bibitem{PPM} H. Sugiyama, K. Nosu ``MPPM: A Method for Improving the Band-utilization Efficiency in Optical PPM.'' {\it Journal of Lightwave Technology}, vol.7, no.3, pp.465-472, 1989.
	\bibitem{Index}  {E. Basar, M. Wen, R. Mesleh, M. D. Renzo, Y. Xiao, and H. Haas, ``Index Modulation Techniques for Next-Generation Wireless Networks,''} {\it IEEE Access}, vol. 5,  pp. 16693-16746, Aug. 2017.
	\bibitem{ImCode2} {G. Kaddoum, Y. Nijsure, and H. Tran, ``Generalized code index modulation technique for high-data-rate communication systems'',} {\it IEEE Trans. Veh. Technol.,} vol. 65, no. 9, pp. 7000-7009, Sept. 2016.
			\bibitem{A2} {X. Chen, T. Chen, and D. Guo, ``Capacity of Gaussian Many-Access Channels,''} {\it IEEE Trans. Inform. Theory}, vol. 63. no. 6, pp. 3516-3539, 2017.
	\bibitem{AA} W. Zhang and L. Huang, ``On OR many-access channels,'' {\it in Proc. IEEE Int. Symp. on Inform. Theory (ISIT)}, Aachen, Germany, June 2017.
	\bibitem{CRC} G. Castagnoli, S. Br\"{a}uer, and M. Herrmann,  ``Optimization of Cyclic Redundancy-Check Codes with 24 and 32 Parity Bits,'' {\it IEEE Trans. on Commun.}, vol. 41, no. 6, pp. 883-892, 1993.
	\bibitem{Cai} T. T. Cai and J. Tiefeng, ``Phase Transition in Limiting Distributions of Coherence of High-dimensional Random Matrices,'' {\it Journal of Multivariate Analysis,} vol. 107, pp. 24-39, 2012.
	\bibitem{gomp} J. Wang, S. Kwon, and B. Shim, ``Generalized Orthogonal Matching Pursuit,'' {\it IEEE Trans. on Sig. proc.,} vol. 60, no. 12, pp.6202-6216, 2012
	\bibitem{mmp} S. Kwon, J. Wang and B. Shim, ``Multipath Matching Pursuit,'' {\it IEEE Trans. Inform. Theory}, vol. 60, no. 5, pp. 2986-3001, May 2014.
	\bibitem{coding} A. J. Viterbi, and K. O. Omura. {\it Principles of digital communication and coding,} Courier Corporation, 2013.
	\bibitem{Joel} J. A. Tropp and A. C. Gilbert, ``Signal Recovery from Random Measurements via Orthogonal Matching Pursuit,'' {\it IEEE Trans. Inform. Theory}, vol. 53, no. 12, pp. 4655-4666, 2007.
	\bibitem{sorm} S. D. Howard, A. R. Calderbank, and S. J. Searle,``A Fast Reconstruction Algorithm for Deterministic Compressive Sensing using Second Order Reed-Muller Codes,'' {\it Proc. of IEEE Conf. on Information Sciences and Systems}, pp. 11-15, 2008.
	\bibitem{wbook} D. Tse and P. Viswanath, {\it Fundamentals of Wireless Communication}, Cambridge University Press, 2005.
	\bibitem{polar} H. Vangala, Y. Hong, and E. Viterbo, ``Efficient Algorithms for Systematic Polar Encoding,'' {\it IEEE Commun. Lett.}, vol. 20, no. 1, pp. 17-20, Jan. 2016.
	\bibitem{book} K. Venkatarama, {\it Probability and Random Processes}, John Wiley \& Sons., 2006.
	\bibitem{A1} {R. Xie, H. Yin, X. Chen, and Z. Wang, ``Many Access for Small Packets Based on Precoding and Sparsity-Aware Recovery,''} {\it IEEE Trans. Commun.}, vol. 64. no. 11, pp. 4680-4694, 2016.

	\bibitem{LDPC} T. J. Richardson and T. L. Urbanke, ``Efficient Encoding of Low-Density Parity-Check Codes,'' {\it IEEE Trans. Inform. Theory,} vol. 47, no.2, Feb. 2001.
	\bibitem{bound}	Y. H. Polyanskiy, P. Vincent, and V. Sergio, ``Channel Coding Rate in the Finite Blocklength Regime,'' {\it IEEE Trans. Inform. Theory,} vol. 56, no. 5, pp.  2307-2359, 2010.

\end{thebibliography}
\end{document}